\documentclass[11pt,letterpaper]{article}
\pdfoutput=1
\usepackage{jheppub}
\usepackage{amsfonts}
\usepackage{amssymb}
\usepackage{graphicx}
\usepackage{latexsym}
\usepackage{amsmath}
\usepackage{float}
\usepackage{etoolbox}

\usepackage{color}
\def\ee{\end{eqnarray}}

\newcommand{\dd}[2]{\frac{\partial #1}{ \partial #2}}

\newcommand{\di}{\partial}

\newcommand{\be}{\begin{eqnarray}}
\newcommand{\en}{\end{eqnarray}}
\newcommand{\bea}[1]{\left(\begin{array}{#1}}
\newcommand{\ena}{\end{array}\right)}
\newcommand{\ba}{\begin{eqnarray}}
\newcommand{\ea}{\end{eqnarray}}

\newcommand{\CO}{\mathcal{O}}

\usepackage{hyperref}

\newcommand\lrpar{\raise .8ex\hbox{$^\leftrightarrow$} \hspace{-9pt}
\partial}
\newcommand\lpar{\raise .8ex\hbox{$^\leftarrow$} \hspace{-9pt}
\partial}
\newcommand\rpar{\raise .8ex\hbox{$^\rightarrow$} \hspace{-9pt}
\partial}

\makeatletter
\def\@fpheader{\relax}
\makeatother

\begin{document} 

\subheader{\ }
\title{ Solving 2D QCD with an adjoint fermion analytically}
\author[]{Emanuel Katz,}
\author[]{Gustavo Marques Tavares,}
\author[]{and Yiming Xu}

\affiliation[]{Physics Department, Boston University, Boston MA 02215}

\date{\today}

\abstract{We present an analytic approach to solving 1+1 dimensional QCD with an adjoint Majorana fermion.  In the UV this theory is described
by a trivial CFT containing free fermions.  The quasi-primary operators of this CFT lead to a discrete basis of states which is useful for diagonalizing
the Hamiltonian of the full strongly interacting theory.  Working at large-$N$, we find that the decoupling of high scaling-dimension quasi-primary operators 
from the low-energy spectrum occurs exponentially fast in their scaling-dimension.   This suggests a scheme, whereby, truncating
the basis to operators of dimension below $\Delta_{max}$, one can calculate the low-energy spectrum, parametrically to an accuracy
of $e^{-\Delta_{max}}$ (although the precise accuracy depends on the state).  Choosing $\Delta_{max} =9.5$ we find very 
good agreement with the known spectrum obtained earlier by numerical DLCQ methods.  Specifically, below the first three-particle threshold, 
we are able to identify all six single-particle bound-states, as well as several two-particle thresholds.}


\maketitle

\section{Introduction}

In this paper we explore a new technique for solving a strongly coupled field theory.  The idea is based on intuition gained from holographic models of strong dynamics where a CFT is broken in the IR by a 
single relevant operator.  The resulting IR states are thus characterized in terms of a single dynamical scale, $\Lambda_s$.  The basic observation, suggested by holography, is
that high scaling-dimension operators (or more precisely, operators in the CFT with a large conformal Casimir) do not have much overlap with the lightest states of the energy spectrum \cite{Fitzpatrick:2013twa}.
In fact, under generic conditions, the analysis of \cite{Fitzpatrick:2013twa} implies that high scaling-dimension operators can decouple exponentially fast from the lightest states. In other words, given
a primary operator of dimension $\Delta$, $\CO(\lambda x) = \lambda^{-\Delta} \CO(x)$, the amplitude of that operator to create the lightest state depends exponentially on its 
dimension: $\langle \Omega |  \CO(0) |\psi \rangle \sim e^{-c \Delta^p}$.  A simplistic bulk model further links the power $p$ to the asymptotic behavior of the density of states, 
predicting that for a QCD-like theory, $p=1$.  Having called the low-dimension sector of a CFT, the effective conformal sector, we will call this conjectured phenomena effective
conformal dominance.  An optimistic interpretation of this dominance is that if one focuses on the lightest states of the theory, then a bulk model which contains
only fields dual to the effective conformal operators might be sufficient.  In particular, if the decoupling of operators occurs exponentially fast, then such models might be useful
even for broken CFTs which do not have a large gap in the scaling-dimensions of operators, such as QCD.  In this work we will find further evidence for this decoupling in a theory whose UV description
is a trivial 2D CFT, that of free fermions in the adjoint of $SU(N)$.  Such a CFT has no gap in the scaling-dimension of operators, and in this sense is quite similar to 4D QCD in the UV.  
However, when we couple the fermions to an $SU(N)$ gauge field through a relevant coupling, we find that the high-scaling dimension operators decouple from the low energy spectrum of 
bound-states exponentially fast.\footnote{This decoupling was observed previously for 2D QCD with fundamental quarks (the 't Hooft model) in~\cite{Katz:2007br}.}
In fact, we will use this observation to propose a new way of solving this theory as an expansion in the maximum scaling-dimension,
$\Delta_{max}$, of the quasi-primary operators used to describe these bound-states.   One may think of this in the holographic language as including only fields in the bulk whose
bulk mass is smaller than a certain number.  Although, the motivation for this approach comes from holographic models, the method we will describe is entirely field theoretic.

\begin{figure}
\begin{center}
\includegraphics[width=0.7\textwidth]{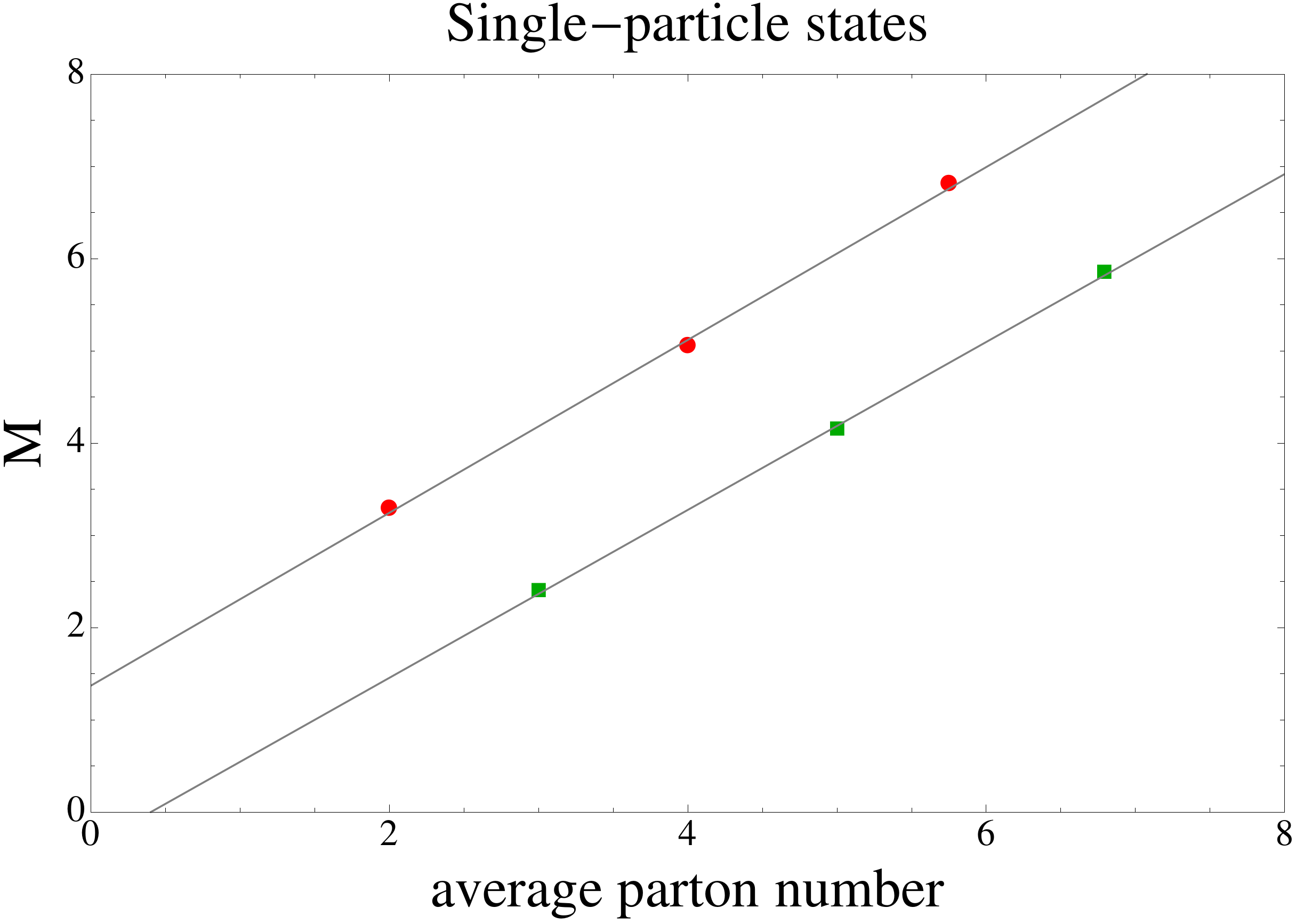}
\end{center}
\caption{The single particle spectrum of adjoint QCD$_{2A}$. }
\label{fig:singleparticle-spectrum}
\end{figure}

We chose QCD$_2$ with an adjoint (massless) Majorana fermion (QCD$_{2A}$) as a laboratory for effective conformal dominance due to its partial similarity to real QCD.\footnote{For an interesting recent application of
a related theory to condensed matter, see \cite{Gopakumar:2012gd}}  As mentioned, the UV of this theory is a very simple
CFT, while its IR contains bound-states of the adjoint quarks.  We will study this theory in the limit of large number of colors, $N$, where many very nice results have been found using 
numerical Discrete Light-cone Quantization (DLCQ) techniques, including a detailed understanding of the low energy spectrum \cite{Dalley:1992yy, Bhanot:1993xp, Gross:1997mx,Trittmann:2001dk}.  Our methods can also be applied away from the large-$N$ limit, however the resulting
spectrum will be more difficult to interpret as most bound-states will develop widths.  Unlike the case of QCD$_2$ with fundamental fermions (i.e. the 't Hooft model \cite{'tHooft:1974hx}), the large-$N$
limit of the adjoint theory does not reduce to quantum mechanics, and thus cannot be be solved analytically using previous techniques.   Indeed, at large-$N$, planar diagrams allow for
adjoint-quark changing processes and therefore generic eigenstates of the Hamiltonian will not be states of definite particle number (however, as has been found numerically,
the single-particle states are largely states of definite particle number).  In this sense this theory is more like real QCD at large-$N$.  Where this theory differs 
from QCD$_4$ is in the behavior of the flux string at  large-$N$.  At large-$N$, QCD$_4$ confines, whereas QCD$_{2A}$ is in the screening phase (in the massless case) \cite{Kutasov:1993gq, Kutasov:1994xq, Gross:1995bp}.
As a result, the spectrum of adjoint QCD$_{2A}$ is quite different.  In particular, the presence of a screening length (which is of order the strong coupling scale), implies
that long strings fall apart, and hence the theory does not contain a Regge trajectory of single-particle states.  Instead, single-particle states can be qualitatively described as
tight bound states of fermions with a definite particle number connected by ``short'' pieces of flux (of about the screening length).  With each additional fermion, one adds another 
``short'' piece of flux to the bound state, leading to the expectation that the mass of the bound states grows linearly with particle number.  This linear growth can be seen readily
in the single-particle spectrum (Fig. \ref{fig:singleparticle-spectrum}).  Since there are very few single-particle states at large-$N$, a natural question is, which states do the many remaining single-trace fermonic operators create?
As argued in \cite{Kutasov:1994xq,Gross:1995bp} and then verified numerically in \cite{Gross:1997mx}, single-trace operators can, surprisingly, create multi-particle states.  The full spectrum at large-$N$ thus contains
both single-particle states as well as multi-particle thresholds.

Our goal in this work will be to reproduce the low-energy spectrum of large-$N$ QCD$_{2A}$, previously obtained numerically via DLCQ.  
Our method is, very roughly, an expansion up to order $e^{-\Delta_{max}}$ for the low-lying states.  For a given $\Delta_{max}$, we expect the expansion to be ineffective for sufficiently excited states.
In practice, we will be approximating the parton wavefunction of a given energy eigenstate, $\Psi(x_1,x_2, x_3, \ ... )$, which is not an eigenstate of particle number, in terms of a basis of
states related to the quasi-primary operators of the free quark theory.  We will see that each quasi-primary operator is described in terms of a particular polynomial in the parton variables, $x_i$,
defined on the simplex spanned by these variables.  Restricting $\Delta_{max}$ is akin to placing a bound on the degree of the polynomials used to approximate the
parton wavefunction.  Hence, in this case, effective conformal dominance is simply the observation that the low-energy states can be very well 
approximated by low-degree polynomials.  The higher the energy of the state, the higher the degree of the polynomial needed to describe it, and therefore
the larger the overlap of the state with a high scaling-dimension operator.  Quasi-primary operators, being orthogonal to each other in the CFT, naturally form
a convenient orthogonal basis of polynomials on a simplex.

We will consider operators up to dimension,  $\Delta_{max} = 9.5$.  This is simply because, to calculate the low-energy spectrum to a high degree of precision, we need 
to evaluate certain multi-dimensional integrals.  Analytic expressions for these integrals can be written down in principle.  However, much like with Feynman diagrams at high-order in perturbation theory, this
quickly becomes very tedious, as the integrals become more complicated, and their number grows exponentially with $\Delta_{max}$.   Instead, we will evaluate these integrals numerically.\footnote{We would like to stress, that if less precision is required, then simple closed form expressions can be provided for the masses
of the lowest single-particle states (see sec. \ref{sing-part} for these).}
These integrals become more time-consuming to evaluate numerically as $\Delta_{max}$ increases, and we have not looked for a particularly efficient 
method for performing this numerical calculation.  
At $\Delta_{max} = 9.5$ our basis contains 810 states when all sectors are considered (as compared to the more than 6700 states typical in a DLCQ calculation).  Still, we will see that though our
basis is much smaller, due to effective conformal dominance, we readily converge to the known results for the first six single-particle states.  We also find
evidence for two-particle thresholds, both in the sectors of what would nominally be the bosonic and the fermionic bound-states (i.e. ignoring the topological sector's
influence on the quantum numbers of a state).  Of course, with a finite basis of states, we do not see a continuous spectrum near the first two-particle 
threshold.  However, knowing the masses of the single-particle states, we are able to properly model the approach to the continuum at a given $\Delta_{max}$
with our method, which we explain in detail in section \ref{multi-part}.  As we will show, we find evidence for the two-particle thresholds associated with the two
lightest fermions.  However we do not, at our $\Delta_{max}$, yet find clear evidence for any two-particle state containing bosons (in either sector).  
We have not attempted to model the approach to the first three-particle threshold in our method, and thus we cannot hope to clearly identify any states
above this threshold.

The paper is organized as follows.   In section \ref{QCD2A} we briefly review QCD$_{2A}$, and its formulation in light-cone coordinates and in light-cone gauge.
This section contains the $M^2$ operator, whose spectrum it will be our task to find.  We then describe our basis, built form conformal quasi-primary operators in section \ref{confops}.  
Next, we present the result of diagonalizing the $M^2$ operator in our basis (truncated at $\Delta_{max} = 9.5$) in section \ref{results}.  We comment on both the 
single-particle spectrum, as well as on the approach to the two-particle continuum.  Finally, we conclude in section  \ref{conclusion}, and make some comments
on the possible lessons of this analysis for holographic modeling.  Some 
technical results are included in the appendices for the benefit of the reader.

\section{Review of QCD$_{2A}$}
\label{QCD2A} 

Two-dimensional QCD with a massless, adjoint quark is most conveniently expressed using light-cone coordinates,  $x^{\pm} = (x^0 \pm x^1)/\sqrt{2}$, and in light-cone gauge $A_- = 0$.   
The action is given by 
\be
S = \int dx^+ dx^- \text{Tr}\left( i \psi \partial_+ \psi + i \chi \partial_- \chi + \frac{1}{2 g^2} (\partial_- A_+)^2  + 2 A_+ \psi \psi  \right),
\label{action}
\ee
where $\psi$ and $\chi$ are respectively the left-moving and right-moving fermion (analogous to chiral fermions in 4 dimensions).  The advantage of light-cone gauge, is
that one can choose $x^+$ to be time, making $\chi$ and $A_+$ non-propagating degrees of freedom. Integrating out these fields one finds that the momentum operators are
\be
\begin{split}
P^+ & = \int dx^- \text{Tr} \left( i \psi \partial_- \psi \right), \\
P^- & = \int dx^- \text{Tr} \left(  -2 g^2 \psi^2 \frac{1}{\partial^2_-} \psi^2 \right).
\end{split}
\label{momentum operator}
\ee

The mass-squared operator is then given by $M^2 = 2 P^+ P^- $.    This operator is commonly written in a basis of Fock-space states, with $k$ $\psi$-quarks of definite $p^+$ momenta: 
$ | p_1, p_2, ... , p_k\rangle$.\footnote{Here and in the future we suppress the $+$ label on the light-cone momenta unless there are potential sources of confusion.}  In this basis, it takes the following form at large-$N$ \cite{Bhanot:1993xp}:
\begin{align}
\label{partonM2}
\langle p_1, p_2, ... , p_k & |   2P^+ P^- | \Psi \rangle  = \frac{ g^2 N}{\pi (x_1 +x_2)^2} \int_0^{x_1+x_2} dy ~\psi_k (y, x_1+x_2 - y, x_3, ... , x_k) \nonumber \\
								& + \frac{g^2 N}{\pi}  \int_0^{x_1+x_2} \frac{dy}{(x_1-y)^2} \left[  \psi_k(x_1, x_2, x_3,..., x_k) - \psi_k (y, x_1+x_2 -y,x_3,...,x_k)\right] \nonumber \\
								& + \frac{g^2 N}{\pi} \int_0^{x_1} dy \int_0^{x_1-y} dz ~\psi_{k+2}(y,z,x_1-y-z,x_2,...,x_k)\left[ \frac{1}{(y+z)^2}-\frac{1}{(x_1-y)^2} \right] \nonumber \\
								& + \frac{g^2 N}{\pi} \psi_{k-2}(x_1+x_2+x_3,x_4,...,x_k)\left[ \frac{1}{(x_1+x_2)^2}-\frac{1}{(x_2+x_3)^2} \right] \nonumber  \\
								& \pm \text{cyclic permutations of } (x_1, x_2, ..., x_k) 		
\end{align}
where the parton variables $x_i = p_i/\sum p_j$, and $\psi_k (x_1,x_2,...,x_k) = \langle p_1, p_2, ...,p_k | \Psi \rangle$ is the component of the parton wavefunction with $k$ quarks. The sign in the last term is always positive 
for odd $k$ and alternates with each cyclic permutation for even $k$.   Note, that unlike the case of the 't Hooft model, there are parton-number changing terms in the above operator.  Such terms are
more typical of relativistic strongly-coupled field theories, and are present in real QCD at large-$N$.

In previous work the spectrum of the mass-squared operator was found by discretizing the light-cone momentum, and expressing this operator as a matrix in the space of multi-parton states carrying
a total of $K$ units of light-cone momentum.  This is done by formally compactifying $x^-$.  
As the integer $K$ is taken to be large, the spectrum of the mass-squared matrix will asymptote to the desired continuum physics.  Our approach will be different in
that we will always be working in the continuum.  Rather, we will express the mass-squared operator in terms of a basis of states which is already naturally discrete even in the continuum theory -
that of gauge-singlet, quasi-primary operators of the free quark CFT.  The resulting matrix can then be diagonalized upon truncation of the infinite list of quasi-primary operators to those of scaling-dimension below $\Delta_{max}$.  We view this as an analytic method, as we are not modifying the mass-squared operator itself or the Hilbert space of the theory (nor are we introducing some external parameter), but merely expressing 
the operator of interest in a convenient basis intrinsic to the continuum theory itself.  Of course, for the purpose of computation, we truncate our basis, for, as we will show, the contribution 
of high scaling-dimension operators to low-energy states is highly suppressed.   This truncation is similar to standard weak-coupling perturbation theory, where, for example, the contribution of 
multi-particle asymptotic final states to a process initiated by a two-particle collision is suppressed.\footnote{In our case, the perturbation theory is not in terms of a small coupling.  Instead, it is more
like the case in the chiral-Lagrangian, where the perturbation is roughly in terms of $\frac{1}{4\pi}$ to a power.  Our perturbation is roughly in terms of $\frac{1}{e}$ to a power set by $\Delta_{max}$.}

\section{Conformal quasi-primary operators}
\label{confops}
In order to obtain the spectrum of the theory we first need to choose an orthogonal basis in order to calculate the mass matrix.
Our results will be calculated truncating the matrix by choosing only a finite set of the basis states. This approach will generate reliable results for the lowest mass states 
as long as there is evidence that increasing the number of basis states does not substantially alter the spectrum (although one must define the precise meaning of this
statement for multi-particle states).  

The choice of basis is motivated by the conjecture of effective conformal dominance. We will therefore choose a basis obtained by acting on the vacuum with operators that correspond to the gauge-singlet quasi-primary operators of the theory in the conformal limit. This choice of basis gives us a natural parameter, $\Delta_{max}$,  with which to truncate the matrix.  In other words, we include only states created by quasi-primary 
operators with scaling-dimensions 
$\Delta \leq \Delta_{max}$.  In a CFT, the two-point function of two quasi-primary operators must vanish unless these operators have exactly the same scaling-dimension (as well as other quantum numbers).  This 
further simplifies the basis, as it leads to automatic orthogonality for most basis states.

We will be interested in states created by quasi-primary operators. In a conformal field theory (CFT) such operators play a central role. In particular, all correlation functions can be calculated using conformal symmetry once the correlation functions of quasi-primary operators are known. Quasi-primary operators are defined as operators that transform covariantly under global conformal transformations. In terms of the generators of conformal transformation $K^{-}$ this condition translates to
\be
[K^-, \mathcal{O}_{\Delta,s}(x^-)]=i\left((x^-)^2\partial_-+ (\Delta - s) \, x^- \right)\mathcal{O}_{\Delta,s}(x^-),
\label{quasi-primary-commutator}
\ee
where $\Delta$ is the dimension of the operator and $s$ its spin.  At $x=0$, the quasi-primary operator,  $\mathcal{O}_{\Delta,s}(0)$, is thus annihilated by $K^-$.  Since we are working in the chiral limit, where the fields $\psi$ and $\chi$ decouple, $K^+$ commutes with operators constructed by $\psi(x^-)$. The descendants of the
quasi-primary \footnote{By descendants, here, we mean only in the sense of the global part of the conformal group, and not the full Virasoro algebra.}, are obtained by acting on the quasi-primary 
with derivatives as $(\partial_+)^n (\partial_-)^m \mathcal{O}_{\Delta,s}(0)$.

Our theory is conformal in the UV, where $ g \rightarrow 0$. Hence, we can ignore the gauge fields, and write the generator $K^-$ in terms of the fermion fields
\be
K^- = & -i\int dx^-  (x^-)^2 \psi_{ij} \partial_- \psi_{ji}.
\ee
Since the left ($\chi$) and right movers ($\psi$) are decoupled, we only need to consider quasi-primary operators built from $\psi$. They are the only dynamical fields in the theory, since $x^{+}$ was chosen as the time coordinate. In the conformal limit, we have $\partial_+ \psi = 0$ by the equation of motion, and therefore we can also neglect operators that have $\partial_+$ acting on $\psi$. Thus, we focus on operators of the form
\be
\mathcal{O}_{n+k/2}\equiv \frac{1}{N^{k/2}}\sum_{\sum s_i=n}c_{s_1, s_2, ..., s_k}\text{Tr}\left(\partial^{s_1}_{-}\psi_1\partial^{s_2}_{-}\psi_2...\partial^{s_k}_{-}\psi_k\right),
\label{primaries1}
\ee
where we have suppressed the spin index, since here $s= - \Delta $ , and substituted $\Delta$ with the appropriate dimension. Note that the coefficient $c_{s_1,s_2,...,s_k}$ has a cyclic symmetry,
\be
c_{s_2,s_3,...,s_k,s_1} = (-1)^{k-1} c_{s_1,s_2,s_3,...,s_k} \, ,
\ee
due to the fermi statistics combined with the trace over gauge indices.

The operators $\mathcal{O}$ defined by eq.~(\ref{primaries1}) automatically satisfies the commutation relation  (\ref{quasi-primary-commutator}) between  $K^+$ and a quasi-primary operator. On the other hand, in order to ensure that $\mathcal{O}$ is a quasi-primary, there are constrains on the coefficients $c_{s_1,s_2,...,s_k}$ from the commutation relation with $K^-$. For example, for operators involving two fields, the lowest dimensional quasi-primary operator is given by $\text{Tr}\left( \psi \overleftrightarrow{\partial_-} \psi \right)$, as can be easily seen by using eq.~(\ref{quasi-primary-commutator}).

Descendants of $\mathcal{O}_{n+k/2}$ are of the form $(\partial_-)^m\mathcal{O}_{n+k/2}$.  As we will ultimately be interested only in finding bound-states which are eigenstates of $P^-$, once we know
the amplitude with which $\mathcal{O}_{n+k/2}$ creates the state, the amplitude with which any of its descendants create this state is determined.  Thus, the descendants do not carry any new dynamical information
which is not already captured by the quasi-primaries.  In sec. \ref{basis} we will see that quasi-primaries can be represented by the set of Jacobi Polynomials in the parton variables, $x_i$.  These 
variables obey the constraint $\sum_i x_i =1$, and the Jacobi Polynomials are known to be a complete basis on a simplex.  This makes it clear that the quasi-primaries already form
a complete basis, and the descendants are not needed in addition.

\subsection{Generating the basis}
\label{basis}
The basis is obtained by acting with quasi-primary operators on the vacuum. It is convenient to perform calculations at fixed $x^+$ since this allows us to expand the field $\psi$ in terms of creation and annihilation operators,
\be
\psi_{ij}=\frac{1}{2\sqrt\pi}\int_0^\infty dp^+\left(b_{ij}(p^+)e^{-ip^+x^-}+b^\dag_{ji}(p^+)e^{ip^+x^-}\right),
\label{mode expansion}
\ee
with $b$ and $b^\dagger$ satisfying the anti-commutation relations
\be
\{ b_{ij}(k^+) , \,b^\dag_{kl}(q^+) \} = \delta (k^+ - q^+) \left( \delta_{ik} \delta_{jl} -  \frac{1}{N} \delta_{ij} \delta_{kl} \right).
\ee
These operators create (destroy) partons with a fixed momentum $k^+$. Since bound states are color singlets, we can restrict our analysis to states created by color singlet combinations of $b^\dag_{ij}$. At leading order in $N$ we need to consider only single trace combinations of $b^\dag$'s because the contribution of multi-trace operators is suppressed. Therefore, in the large $N$ limit, a complete set of color singlet states is given by
\be
| p_1 , p_2 , ... , p_k \rangle = \frac{1}{N^{k/2}} \text{Tr}\left( b^\dag (p_1) b^\dag (p_2) ... b^\dag (p_k) \right) |0\rangle.
\label{singletstate}
\ee

The quasi-primary operators of the free fermion CFT naturally have a fixed number of fields, therefore the basis states have a well defined number of partons. In addition, since the Hamiltonian conserves the number of partons of a given state modulo 2~(see Appendix \ref{appmassmatr}), the mass matrix is automatically block diagonal with respect to bound states being fermions (odd number of partons) or bosons (even number of partons).

The Hamiltonian also has a $Z_2$ symmetry, which we denote by T-parity, under which the fields transform as $T \psi_{ij} = \psi_{ji}$. We can choose our quasi-primary operators to have definite T-parity charge, i.e, $T \mathcal{O} = (-1)^T \mathcal{O}$. By doing this we further break the mass matrix into four blocks, $(\text{bosonic or fermionic})\otimes(\text{T even or odd})$.

As discussed in Appendix \ref{appquasi-primary}, the wave-functions of the (Fourier transformed) quasi-primary operators, 
\be
\delta\left( \sum p_i - P \right) \, f(p_1,p_2,...,p_k) = \langle p_1, p_2, ..., p_k |  \mathcal{\tilde O}_{n+k/2} (P) |0\rangle
\label{wavefn1}
\ee
have a special form when expressed in term of angular variables defined by
\be
\begin{split}
p_{k}&=P \cos^2\theta_1,\\
p_{k-1}&=P \sin^2\theta_1 \cos^2\theta_2,\\
&...\\
p_{2}&=P \sin^2\theta_1 \sin^2\theta_2...\cos^2\theta_{k-1},\\
p_1&=P \sin^2\theta_1 \sin^2\theta_2...\sin^2\theta_{k-1}.\\
\end{split}
\nonumber
\ee
In these variables the wave-functions can be expressed as linear combinations of products of Jacobi Polynomials of the form~\footnote{
Recall that the Legendre Polynomial is a special case of Jacobi Polynomials $P_{l} = P^{(0,0)}_{l}$.
}
\be
\begin{split}
&f_{n,l_1,l_2,...,l_{k-2}}\left(P, \theta_1, \theta_2...,\theta_{k-1}\right)\\
&=P^n \sin^{2l_1}\! \theta_1\sin^{2l_2} \! \theta_2 ... \sin^{2l_{k-2}} \! \theta_{k-2}\\
&\times P^{(2l_1+k-2,0)}_{n-l_1}\left(\cos2\theta_1\right)P^{(2l_2+k-3,0)}_{l_1-l_2}\left(\cos2\theta_2\right)...P^{(2l_{k-2}+1,0)}_{l_{k-3}-l_{k-2}}\left(\cos2\theta_{k-2}\right)P_{l_{k-2}}\left(\cos2\theta_{k-1}\right),
\end{split}
\label{wavefnbasis}
\ee
with $n \geq l_1 \geq l_2 \geq ... \geq l_{k-2}$. All wave-functions can be generated by these ``basis'' functions with the additional constraint that they lead to wavefunctions cyclic in the original momenta variables (up to a $(-1)^k$ sign). This constraint comes from the single trace requirement. This reproduces the previous result \cite{Katz:2007br} that for operators bilinear in the fields, the wave-function is given by Legendre Polynomials. 

We also need to ensure that our basis is orthonormal. Operators with different dimensions, different T-parity or different number of partons are automatically orthogonal. However given two operators of same dimension with equal T-parity and number of partons, $\mathcal{O}_i$ and $\mathcal{O}_j$, they must satisfy
\be
\begin{split}
\langle 0 | \mathcal{\tilde O}_i^\dag (P) \mathcal{\tilde O}_j(P^{'}) | 0 \rangle &=  \frac{1}{k} \delta(P-P^{'}) \int_{\sum_i p_i =P} \prod_i dp_i g(p_1,p_2,...,p_k)^* f(p_1,p_2,...,p_k) \\
& = \delta_{ij} \delta(P-P^{'}).
\end{split}
\ee
Here $g$ and $f$ are the corresponding wavefunctions defined in eq. (\ref{wavefn1}). The factor $1/k$ takes care of the normalization of the color singlet state ~(\ref{singletstate}) at leading order in $1/N$.

\begin{table}
\centering
\begin{tabular}{|c|c|c|c|c|c|c|c|c|}\hline $\Delta_{max}$ & 2 & 3 & 4 & 5 & 6 & 7 & 8 & 9 \\\hline T-even & 1 & 1 & 4 & 5 & 16 & 27 & 75 & 153 \\\hline T-odd & 0 & 1 & 2 & 6 & 12 & 31 & 66 & 165 \\\hline \end{tabular}\caption{Number of states in the bosonic sector at each operator dimension}\label{table:bosonstatenum}
\end{table}

\begin{table}
\centering
\begin{tabular}{|c|c|c|c|c|c|c|c|c|c|}\hline $\Delta_{max}$ & 1.5 & 2.5 & 3.5 & 4.5 & 5.5 & 6.5 & 7.5 & 8.5 & 9.5 \\\hline T-even & 0 & 1 & 1 & 5 & 7 & 22 & 42 & 111 & 235 \\\hline T-odd & 1 & 1 & 3 & 4 & 11 & 18 & 51 & 99 & 257 \\\hline\end{tabular}\caption{Number of states in the fermionic sector at each operator dimension} \label{table:fermionstatenum}
\end{table}

The number of states in the basis of conformal quasi-primaries up to dimension 9 for the bosonic sector and 9.5 for the fermionic sector is listed in Tables \ref{table:bosonstatenum} and \ref{table:fermionstatenum}.  Note that the dimension of the basis is almost an order of magnitude smaller than that used for the DLCQ method. 

The matrix element of the mass operator $2P^{+}P^{-}$ in the quasi-primary operator basis is  defined as 
\be
M^2_{i,j}=\langle\mathcal{O}_i|2P^{+}P^{-}|\mathcal{O}_j\rangle.
\ee
For example, there are 5 quasi-primary operators up to $\Delta_{max} =5$ in the T-even sector 
\be
\begin{split}
\mathcal{O}_1&\sim \text{Tr}\left((\partial \psi)\psi-\psi\partial\psi\right) ,\\
\mathcal{O}_2&\sim \text{Tr}\left((\partial^3 \psi)\psi-9(\partial^2\psi)\partial\psi\right)  \pm ... ,\\
\mathcal{O}_3&\sim \text{Tr}\left((\partial \psi)(\partial\psi)\psi\psi\right) \pm ... ,\\
\mathcal{O}_4&\sim \text{Tr}\left((\partial \psi)\psi\psi\psi\psi\psi\right) \pm ... ,\\
\mathcal{O}_5&\sim \text{Tr}\left((\partial^2 \psi)\psi\psi\psi\psi\psi-2 (\partial \psi)\psi(\partial\psi)\psi\psi\psi\right) \pm ... ,\\
\end{split}
\ee
where the ellipses refer to terms related to the first one by cyclic permutations.  The matrix elements are calculated by numerically integrating the wavefunctions (\ref{wavefn1}), obtained directly from the solutions given by 
eq.~(\ref{wavefnbasis}), against the parton basis mass-squared operator of eq.~(\ref{partonM2}). A more detailed explanation of the calculation can be found in the Appendix \ref{appmassmatr}. In the case of the above 5 quasi-primary 
operators, for example, the corresponding $5\times 5$ dimensional mass matrix is
\be
\label{examplematrix}
\left(
\begin{array}{cccccc}
 12. & 3.05 & 4.83 & 0 & 0  \\
 3.05 & 51.3 & -7.38 & 0 & 0  \\
 4.83 & -7.38 & 44.3 & 0 & 0  \\
0 & 0 & 0 & 56. & 0  \\
0 & 0 & 0 & 0 & 72. \\
\end{array}
\right).
\ee

The spectrum of the adjoint fermion model is then obtained from the diagonalization of the mass matrix with a basis up to a maximum operator dimension $\Delta_{max}$.

\section{Results}
\label{results}

In this section we present the spectrum of the mass-squared operator in the conformal quasi-primary basis. We diagonalize the mass-squared operator for a basis of up to dimension 9 for the bosonic sector, and 
up to 9.5 for the fermonic sector. Since there is no mixing between T-even and odd states, the diagonalization is performed separately for each sector. In total we found six single particle states, together with states that match well to the expected manifestation of a continuous spectrum at finite $\Delta_{max}$.   This agrees with previous studies \cite{Dalley:1992yy, Bhanot:1993xp, Gross:1997mx,Trittmann:2001dk}. Most of the single particle states have clearly converged.   That is, for these eigenstates we saw an exponential drop in the contribution of high-dimension operators, with the combined weight carried by operators of order 
$\Delta_{max}$ being at or below $10^{-3}$. Specifically, the lowest states in each sector converge at a rather small dimension when the basis contains only $\sim 50$ states for all sectors combined. 
Therefore, different from DLCQ, we obtain the particle mass without the need to linearly extrapolate. The mass spectrum of the single particle states we find grows linearly with parton number. 

As discovered previously \cite{Gross:1997mx}, we also saw evidence for a continuous spectrum starting with twice the mass of the lightest state $|F_1\rangle$.  We will describe the way one can 
identify a two-particle threshold with a truncated basis using our approach in section \ref{multi-part}.

\begin{figure}
\begin{center}
\includegraphics[width=0.78\textwidth]{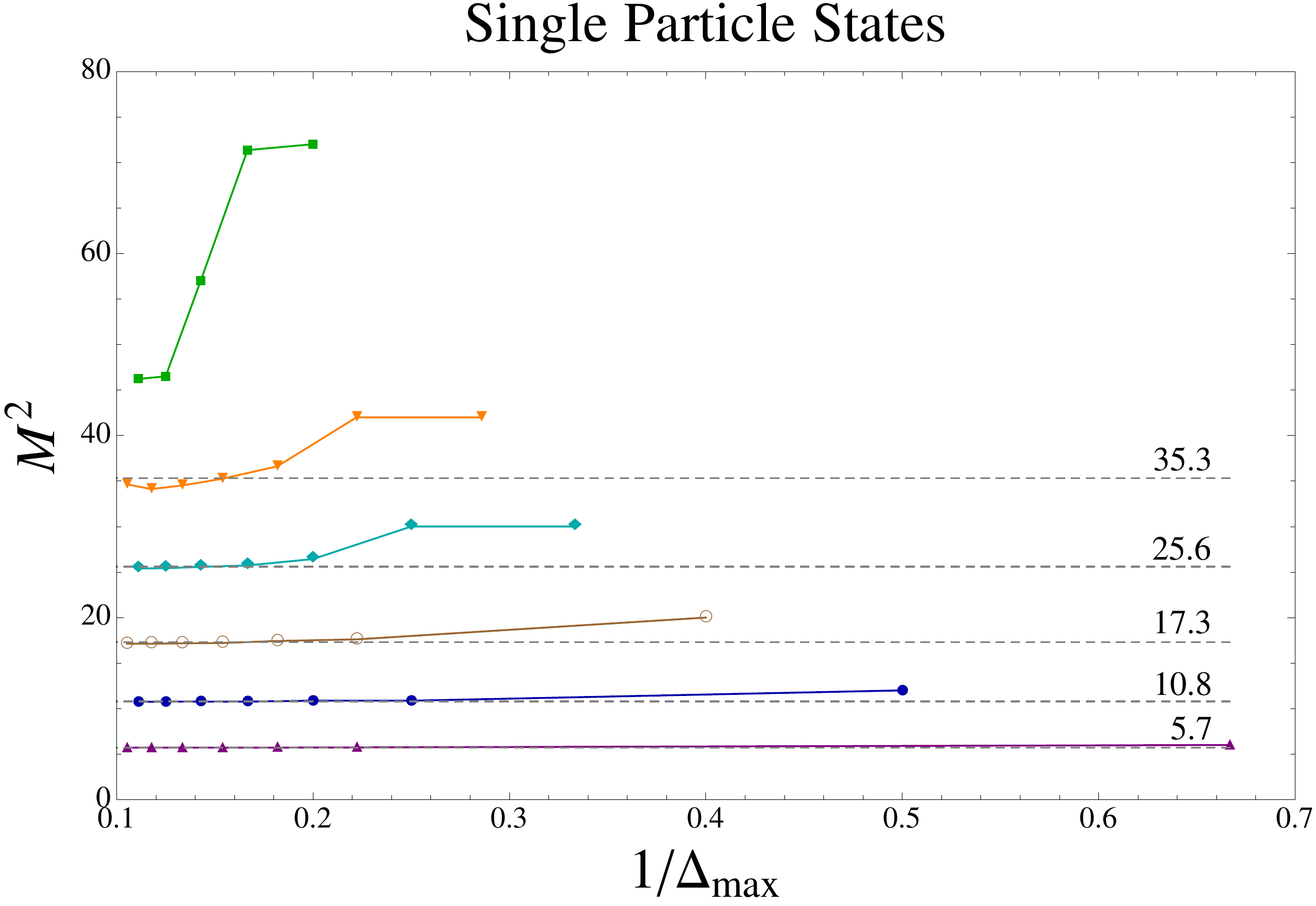}
\includegraphics[width=0.78\textwidth]{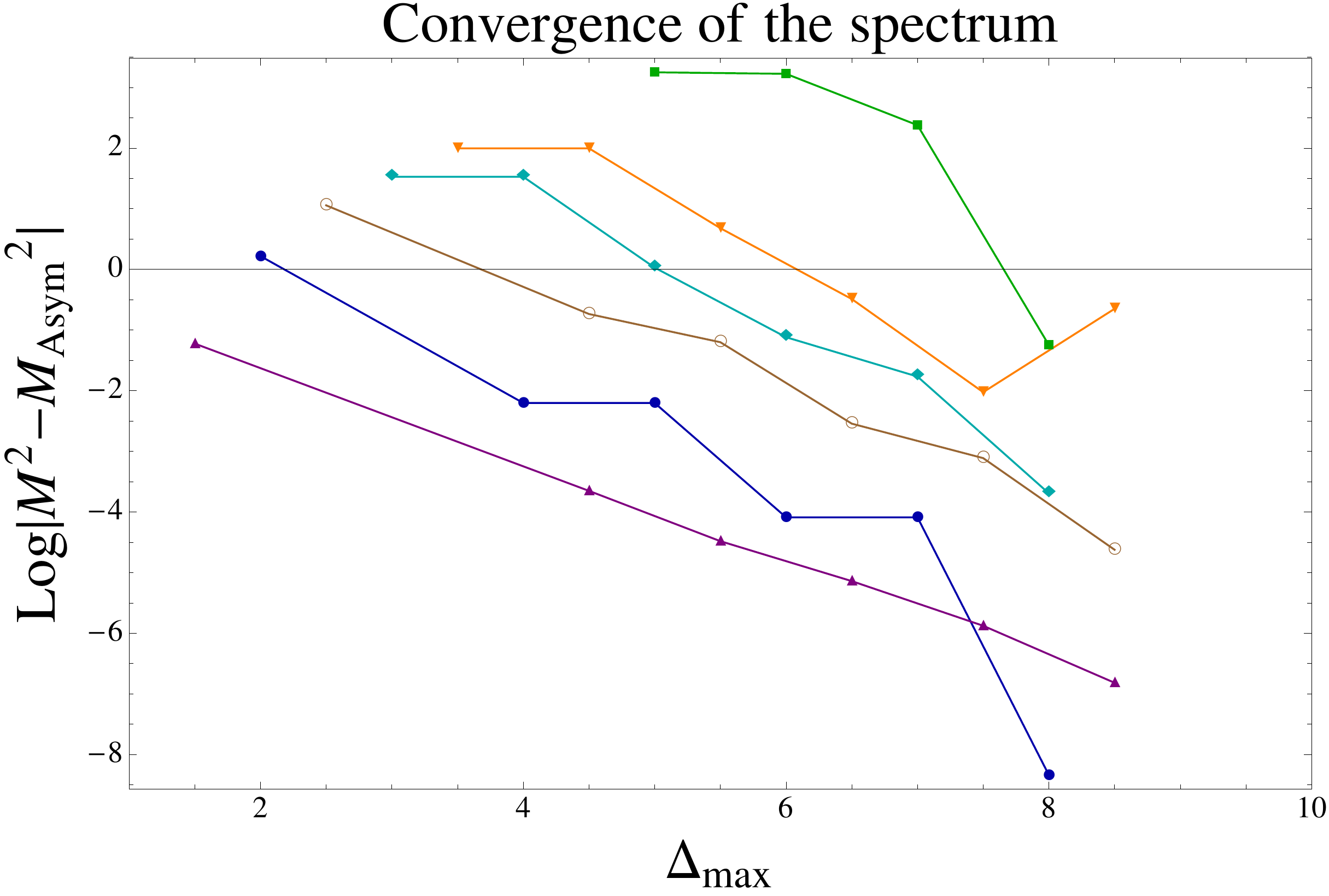}
\end{center}
\caption{The convergence of the spectrum of the low-lying single-particle-states.
Here $\Delta_{max}$ is the dimension of the highest quasi-primary operator used to generate a truncated Hilbert space. In the bosonic sector we calculated the spectrum up to $\Delta_{max}=9$, whereas in the fermionic sector the largest $\Delta_{max}$ is equal to 9.5. The second plot demonstrates more clearly the degree of convergence of the single-particle states.  All states but the highest one appear to have a similar rate of convergence.
The asymptotic masses are taken to be the values at the highest $\Delta_{max}$ calculated. The spectrum appears to converge to the asymptotic values parametrically as $e^{-\Delta_{max}}$.
\label{fig:singleparticle-invdelta-plot}}
\end{figure}

\begin{figure}
\begin{center}
\includegraphics[width=0.9\textwidth]{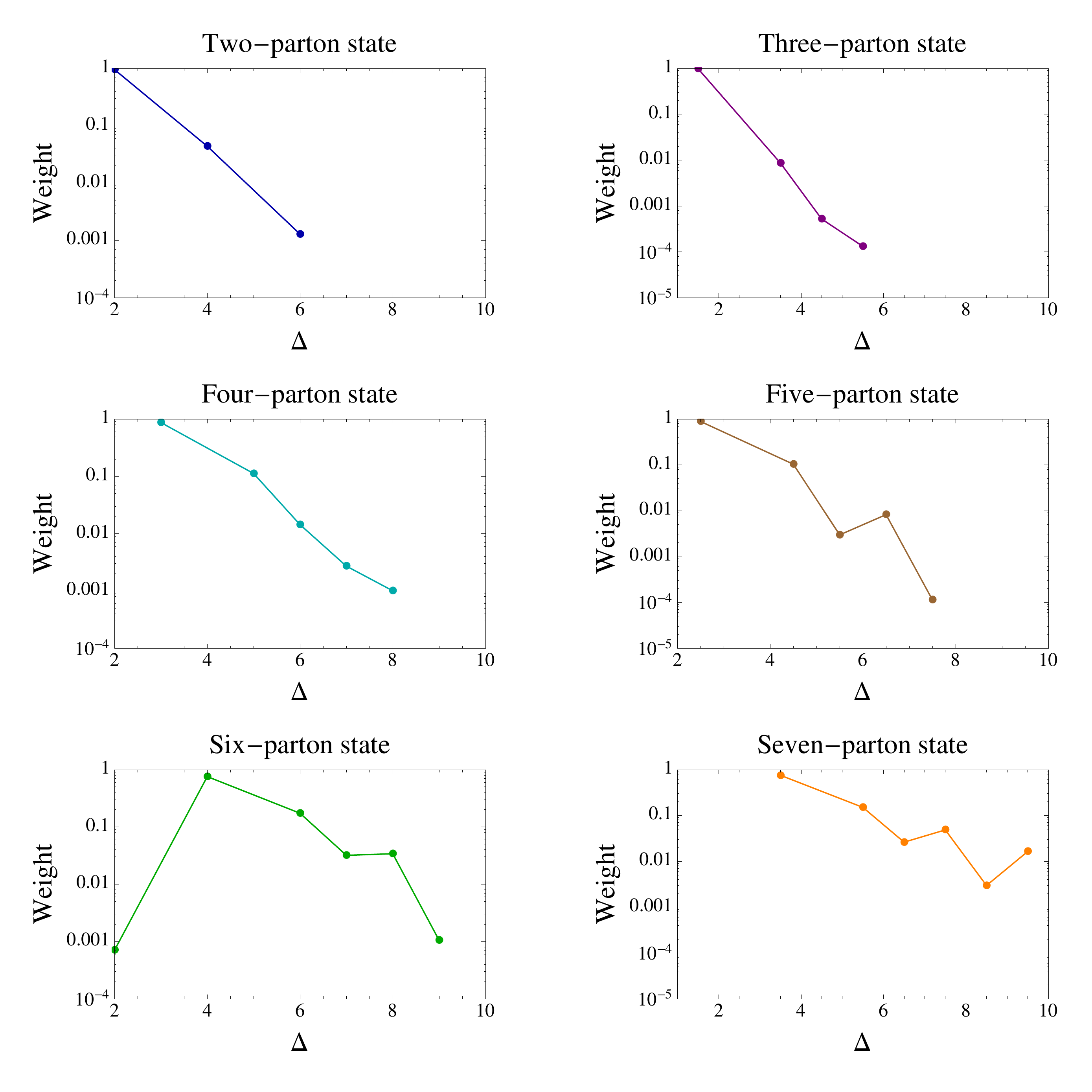}
\end{center}
\caption{The weight in dimension for the six single particle states. 
In the left and the right column are the bosonic states and the fermionic states, respectively. The color code for each state is the same as that in Fig. \ref{fig:singleparticle-invdelta-plot}.
\label{fig:singleparticle-amplitudes}}
\end{figure}

\subsection{Single particle states}
\label{sing-part}
We show in Fig. \ref{fig:singleparticle-invdelta-plot} the convergence of the mass  with respect to $1/\Delta_{max}$ for the eigenstates identified as single particle states. Here $\Delta_{max}$ is equal to the dimension of the highest quasi-primary operator used to generate a truncated Hilbert space. The mass-squared, $m^2$, is in units of $g^2 N/\pi$. States are counted as single particle states if more than $0.9$ of their weight is generated by operators with the same number of partons. It is therefore straightforward to track these states as $\Delta_{max}$ increases. It is remarkable that the spectrum of the low-lying states, especially the ground states, converge rapidly even at low dimension. We obtained a spectrum with the masses in the bosonic sector being $m^2 = 10.7 ~(2),  25.4~(4),  46.2~(6)$\footnote{
The weight of this state generated by 6-parton operators is 0.84. Although it does not meet our criteria for single-particle states at current resolution, our numerical results (see Fig. \ref{fig:singleparticle-invdelta-plot}) suggest that this state has not stabilized in $\Delta_{max}$ compared to the lighter states. In addition to fitting the linear mass prediction, the wave function of this state has qualitative features that are much closer to single-particle states than to multi-particle states. Therefore we identify this state as a single-particle state.
}
, and the fermionic sector being  $m^2 = 5.7~(3), 17.1~(5), 34.6~(7)$, with the numbers in the parentheses showing the corresponding parton number. The dashed lines in the figure indicate the DLCQ results \cite{Gross:1997mx}, showing a good agreement between the two methods.

It is easy to see a linear relation between the single particle mass $m$ and the parton number, in either the bosonic or fermionic sector, as illustrated in Fig.~\ref{fig:singleparticle-spectrum}. The deviation of this linear fit is less than $0.05$. This agrees with previous studies \cite{Bhanot:1993xp}.   Let us make two more comments about this spectrum. First, the bosonic spectrum starts with a heavier ground state than the fermionic sector because of fermion statistics.  
The two-parton wavefunction is antisymmetric under a cyclic permutation of partons, but this is not the case for a three-parton state. Consequently, the three-parton state is to leading order a constant (zero degree polynomial), 
while the two-particle state depends linearly on the parton variables to leading order.  Integrating the approximate polynomial wavefunction against the $M^2$ operator, then results in  the lowest antisymmetric two-parton state being heavier than the three-parton ground state (as we show explicitly below).  Second, regarding, the heaviest single-particle state that we have observed, given the linearity of the spectrum,  and the onset of decoupling (which can be seen in Fig. \ref{fig:singleparticle-amplitudes}), we conjecture that the mass of the six-parton state (green line in Fig.  \ref{fig:singleparticle-invdelta-plot}) should settle at $m^2=46.2$ within a $0.01$ variation, despite the fact that its mass does not start converging until at $\Delta_{max}=8$.

In Fig. \ref{fig:singleparticle-amplitudes} we show the logarithmic plots of the weight at each dimension $\Delta$ for the single particle states.  By weight, we mean the sum of squared-amplitudes of a state 
generated by all operators at dimension $\Delta$, $\sum_\Delta \langle\Omega|\mathcal{O}_\Delta| \psi \rangle^2$. In all sectors the weight drops exponentially.  Except for the seven-parton state, at dimension 9 (9.5) for bosons (fermions)  all the single particle states have reached a point where the variation of the weight at next dimension will be less than  $10^{-3}$. We therefore consider these states decoupled from higher dimension operators. Their mass will not change by more than ${10^{-3}}$ when we enlarge the basis. In particular,  the bosonic T-even lightest state (a two-parton state) and the fermionic T-odd lightest state (a three-parton state) converge so quickly that only a basis with a few states is needed in each sector. The rapid decoupling of these states provides a consistency check on the convergence of the 
mass as shown in Fig. \ref{fig:singleparticle-invdelta-plot}. 

It is also remarkable that the exponential decoupling rates of all decoupled single particle states are similar. The weight at dimension $\Delta$ behaves parametrically as $\exp(-\alpha\Delta)$. Excluding the not-yet-decoupled seven-parton state, all other single particle states have $\alpha \sim 2$, varying  from $1.5$ to $2.3$. In general the change of the slope due to a possible mismatch of the decoupling dimension, where the tail of the exponential falling pattern starts, is only within order of $O(0.1)$ and will not affect the number significantly..

Given the rapid decoupling in our basis, we can easily write down analytic formulae for the ground state wavefunctions, since they are generated by very few operators. The wavefunctions then enable a leading order calculation of the mass. For example, at $\Delta_{max}=9$, the lightest bosonic state has probability of more than $0.96$ of being a two-parton state, and is dominated by the lowest two-fermion operator. At leading order its wavefunction can be parametrized by a single Legendre polynomial, $P_{2-p}(x_1,x_2)\equiv\langle \theta| \mathcal{O}_{1+1}\rangle=\sqrt{6}(1-2\sin^2\theta)=\sqrt{6} (x_2-x_1)$.
Here $x_i$ is the ratio of the momentum of the $i^{\text{th}}$ parton to the total momentum, $p_i/P$.
The numbers in the square-roots account for the wavefunction normalization.
Its mass, up to a $\sim 15\%$ correction (compared to its asymptotic value  $m^2=10.7 g^2N/\pi$), can be obtained from
\be
\begin{split}
&\langle2\text{ - parton} |M^2|2\text{ - parton}\rangle\\
&=\frac{g^2N}{\pi}\int^1_0 dx_1dx_2\delta(x_1+x_2-1) \int^1_0 dy\frac{6 \left((x_2-x_1) - (1- 2 y) \right)^2 }{2(x_1-y)^2}\\
&=12 \times\frac{g^2N}{\pi}.
\end{split}
\ee
Note that this just corresponds to the first diagonal entry of the matrix in eq.(\ref{examplematrix}). One can include higher order corrections by diagonalizing the full $3\times3$ upper block of this matrix, which includes the contribution of operators up to dimension 4. This reduces the error in the mass to less than 2$\%$.  The corresponding parton wavefunction would be given by a slightly more lengthy expression, containing both a 2-parton, and a 4-parton component, but it is straightforward to write down.

As another example, the wavefunction of a three-parton ground state is with $0.99$ weight captured by the operator $\mathcal{O}_{3/2}=\text{Tr}\left({\psi_1\psi_2\psi_3}\right)$.  Thus, its mass can be simply given by
\be
\begin{split}
&\langle 3\text{ - parton} |M^2|3\text{ - parton}\rangle\\
&=\frac{g^2N}{\pi}\int dx_1dx_2dx_3\delta\left(x_1+x_2+x_3-1\right)(\sqrt{6})^2\frac{1}{(x_1+x_2)^2}\int^{x_1+x_2}_0 dy\\
&=6\times\frac{g^2N}{\pi},
\end{split}
\ee
where $\sqrt{6}$ is a wavefunction normalization. The cyclic symmetry of the wavefunction and the mass operator is also taken into account in this formula. This leading order result differs from the real mass  $m^2=5.7 g^2N/\pi$ by only $5\%$.

\subsection{Multi-particle states and the continuous spectrum}
\label{multi-part}

\begin{figure}[h]
\begin{center}
\includegraphics[width=0.8\textwidth]{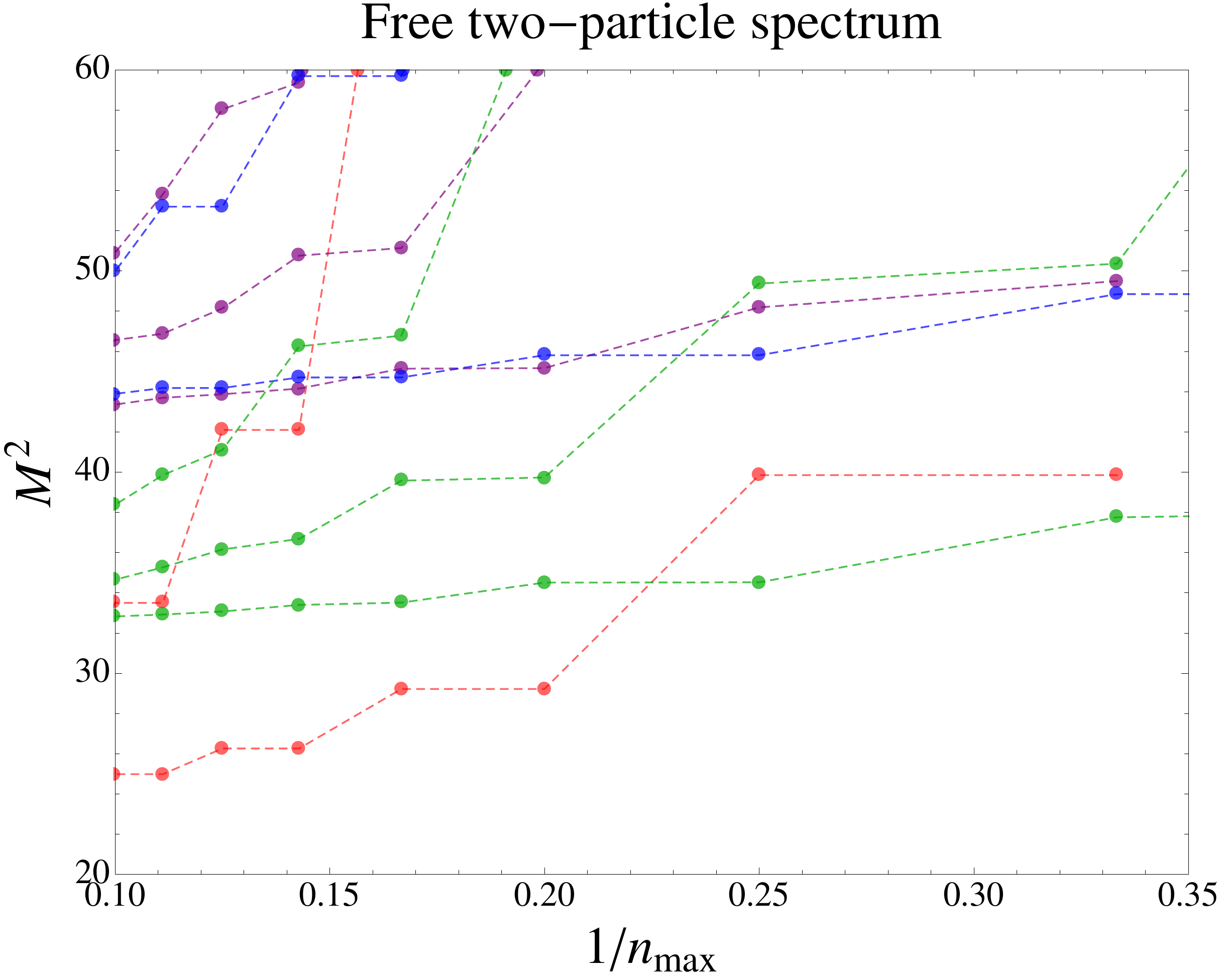}
\end{center}
\caption{ The convergence of the free two-particle spectrum, as a function of $1/n_{max}$ to the continuum. In the truncated basis, $n_{max}$ is the largest degree of Jacobi polynomials $P_n^{(a,b)}$ used, corresponding to quasi-primary operators below a certain maximum dimension. The construction of the operator basis can be found in  Appendix \ref{freetwopartbasis}. The red, purple, green and blue lines plot the expected spectra of free two-particle states $F_1\otimes F_1$, $F_1\otimes F_2$, $F_1\otimes B_1$ and $B_1\otimes B_1$, respectively. 
\label{fig:freetwopartspectrum}}
\end{figure}

\begin{figure}[h]
\includegraphics[width=0.5\textwidth]{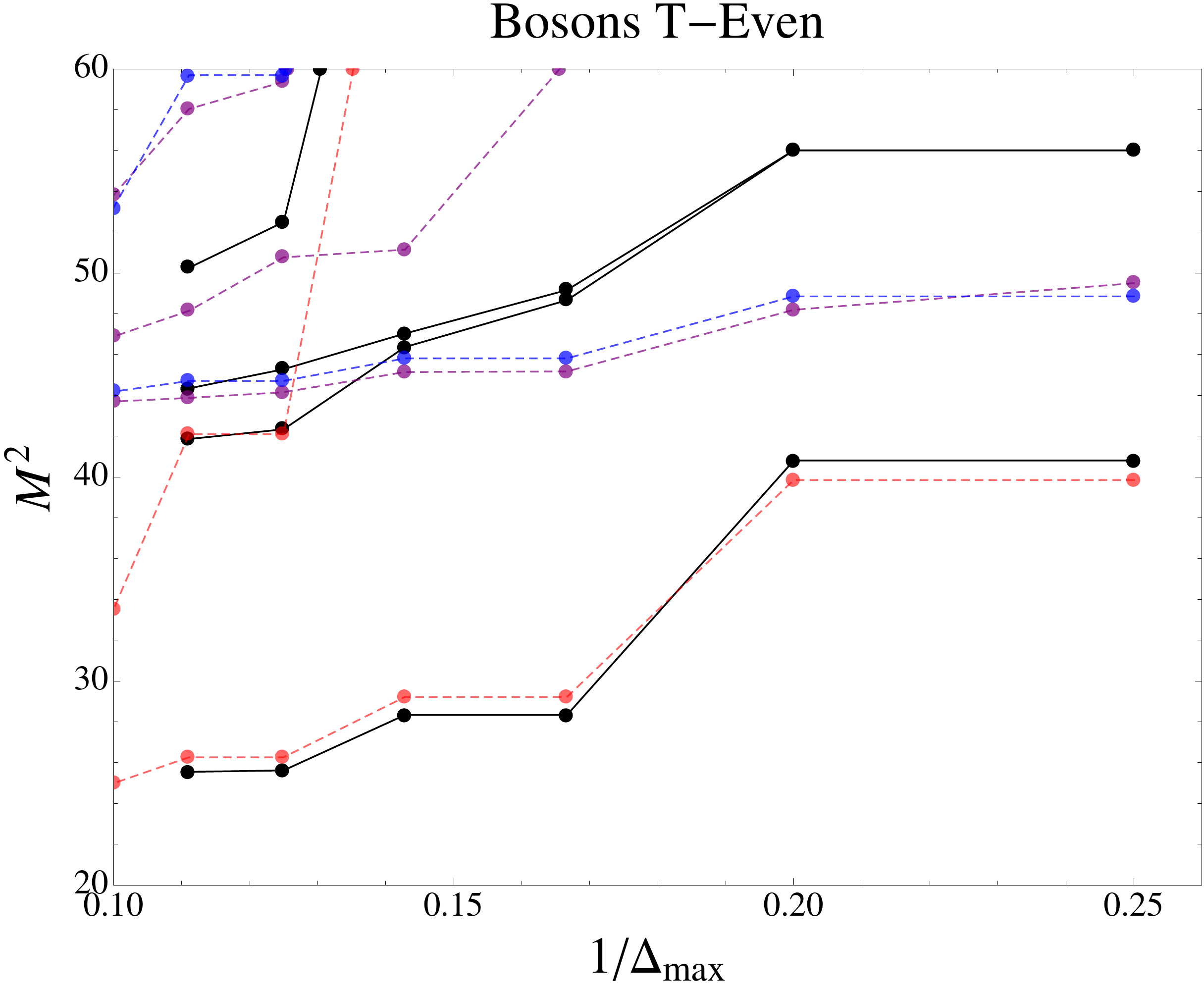}
\includegraphics[width=0.5\textwidth]{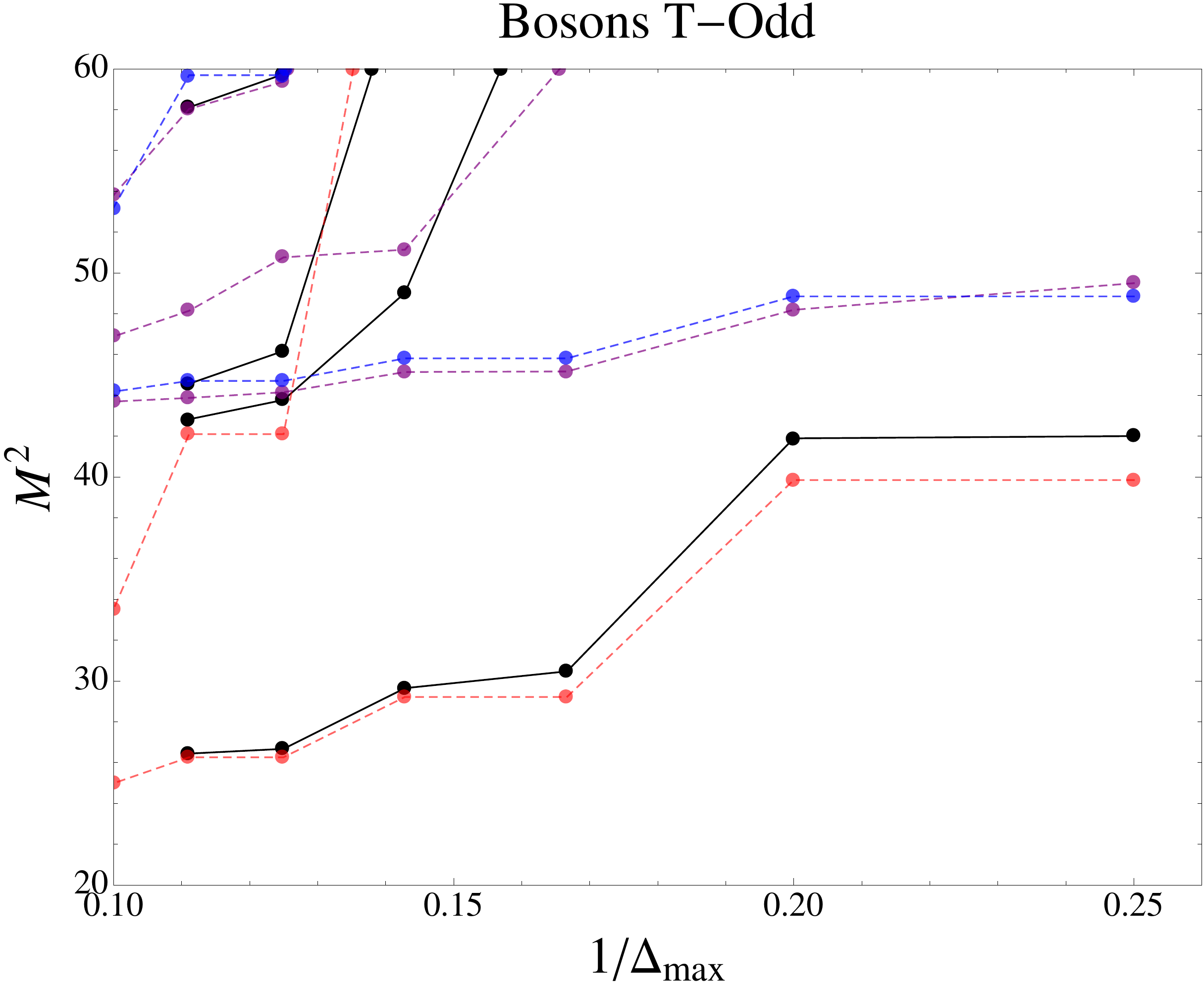}
\includegraphics[width=0.5\textwidth]{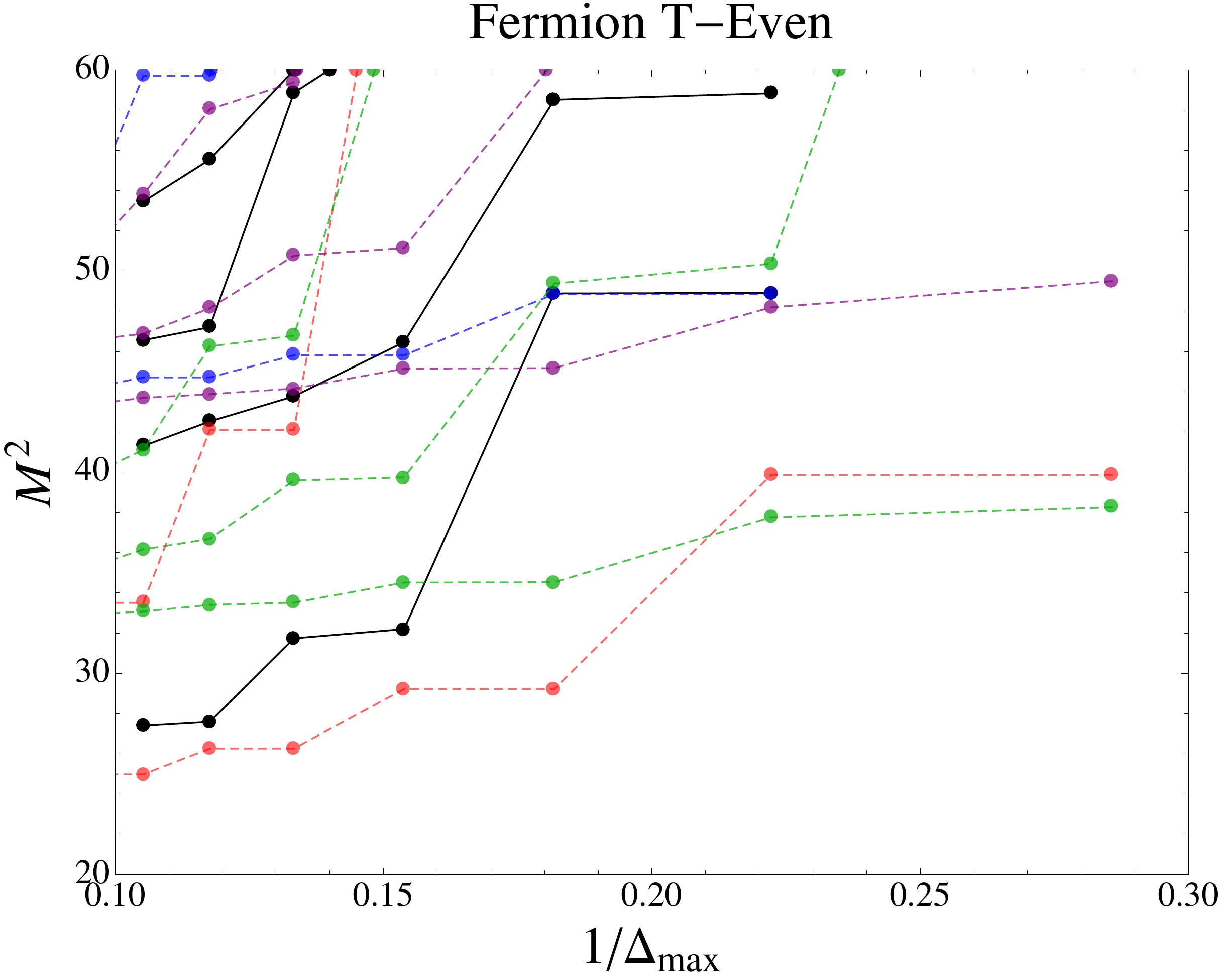}
\includegraphics[width=0.5\textwidth]{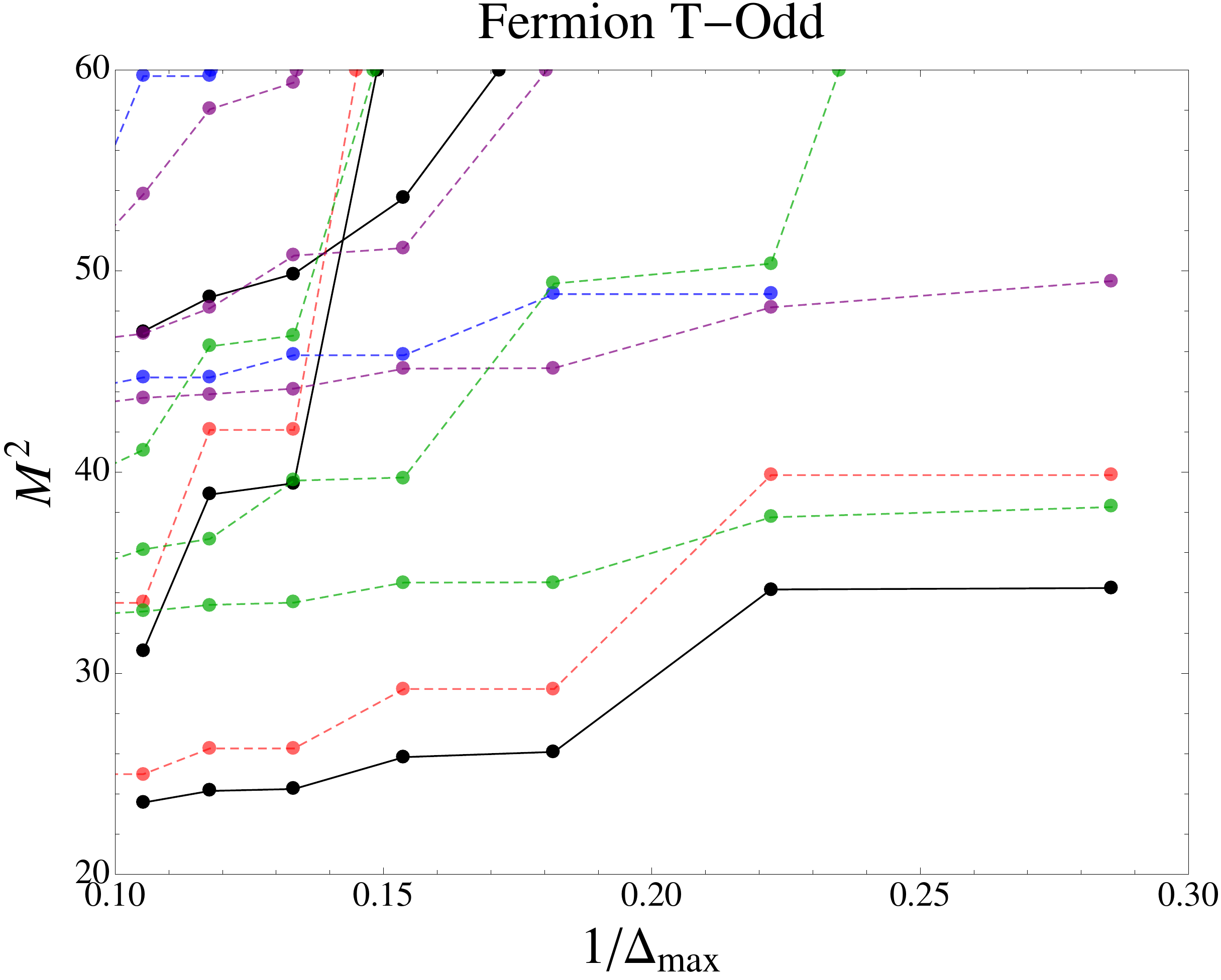}
\caption{The spectra of the multi-particle states in the four sectors with a given $T$-parity and statistics. They are compared with the free two-particle spectra of states $F_1\otimes F_1$ (red), $F_1\otimes F_2$ (purple), $F_1\otimes B_1$ (green) and $B_1\otimes B_1$ (blue). The $F_1\otimes B_1$ spectrum is not included in the bosonic sectors because there is  no obvious counterpart of this state in the QCD$_{2A}$ spectrum.
\label{fig:multipartspectrum} }
\end{figure}

How should we interpret the remaining states, not included in the discussion above?  These states are not approximate eigenstates of particle number.  From earlier work, these must be somehow related to the continuum, beginning at the first two-particle threshold, $m=2m_{F_1}$, twice the mass of the ground state fermion.   Thus, we must understand how our truncated, finite, Hilbert space is trying to approximate a continuum.

In this work, we will only study the approach to the two-particle continuum, and therefore our task will be to understand the spectrum below $(3 m_{F_1})^2 = 51.5$, the first three-particle threshold.  The $M^2$ spectrum
of a free two-particle state is given by
\be
M_{2part}^2(x)=\frac{m_1^2}{x}+\frac{m_2^2}{1-x}.
\ee
Here, $m_i$ are the masses of the individual particles, and $x$ labels the momentum fraction of one of these particles.  How do we see such a spectrum at finite $\Delta_{max}$?  Let us assume
that diagonalizing the full $M^2$ operator, we have converged to this spectrum, in some range of energies.  In this case, the states we have found can be modeled by free two-particle states of a particular
$M_{2part}^2(x)$, with the masses $m_1$ and $m_2$ chosen from the spectrum of single-particle states obtained earlier, and we are interested in plotting the spectrum as a function of $\Delta_{max}$.  In the case where $m_1$ and $m_2$ correspond to single-particle states in the fermionic sector, the quasi-primary operators which create the two-particle state, of two free particles are of the form
\be
\mathcal{O}^{2-part}_{\Delta}\sim\sum_k\left(\frac{n!}{k!(n-k)!}\right)^2 \partial^k \Psi_1 \partial^{n-k}\Psi_2.
\ee
The general form of the quasi-primary operators, when one or two of the masses correspond to state in the bosonic sector, is described in Appendix~\ref{freetwopartbasis}. Here, $\Psi_i$ are are free fields which create particles of masses $m_i$, and $\Delta=n+1$ for two fermions, $\Delta=n+\frac{1}{2}$ for a boson and a fermion and $\Delta = n$ for a two bosons.
In terms of these operators, we can construct a basis of states, and calculate the $M_{2part}^2(x)$ matrix in this basis:
\be
[M_{2part}^2]_{\Delta,\Delta'}=\int^1_0 dx \phi^{*}_\Delta(x)\left(\frac{m_1^2}{x}+\frac{m_2^2}{1-x}\right)\phi_{\Delta'}(x),
\label{freetheorymass}
\ee
where $\phi_\Delta(x)\equiv \langle x, 1-x|\tilde{\mathcal{O}}^{2-part}_{\Delta}|0\rangle$, $\tilde{\mathcal{O}}^{2-part}_{\Delta}$ defined as the Fourier transform of the quasi-primary operator, with $x$ being the momentum
fraction of one of the particles.\footnote{
In practice the integral in eq.~(\ref{freetheorymass}) has divergences as $x\rightarrow 0$ and $x\rightarrow 1$. We deal with these divergences by integrating from $\epsilon$ to $1-\epsilon$ and taking $\epsilon$ exponentially small. One can easily show that this cutoff does not affect the smallest eigenvalues of the matrix, which are the ones we use for a comparison with the full theory calculation.
}
As mentioned in section \ref{confops}, if both masses are associated with single-particles states in the fermionic sector, $\phi_\Delta(x)$ will simply be a Legendre polynomial of a degree $n$ related to $\Delta$ as above.  Note, that if both particles are
identical fermions then only the antisymmetric states in the basis are allowed, in other words, only odd-degree Legendre polynomials occur.  In Fig. \ref{fig:freetwopartspectrum}  we provide a plot of the result of diagonalizing
$[M_{2part}^2]_{\Delta,\Delta'}$ for $\Delta\leq \Delta_{max}$, as a function of $ \Delta_{max}$.  Different color trajectories towards the continuum correspond to the the different two-particle states expected below the
first three-particle threshold, where we have put in the appropriate masses $m_i$ obtained from the lowest three single-particle states ($|F_1\rangle$, $|B_1\rangle$, and $|F_2\rangle$) into eq.~(\ref{freetheorymass}).  Note, that the trajectories associated with $|F_1\rangle \otimes |F_1\rangle$ (the red-dashed lines) are more sparse due to the fermionic statistics, as discussed.
 
Having understood the expected properties of the continuum spectrum for finite $\Delta_{max}$, we can now compare this to our results from diagonalizing the full $M^2$ as a function of $\Delta_{max}$.
The results are presented in Fig. \ref{fig:multipartspectrum}   where we show separately the four sectors with a given T-parity and nominal statistics.   The agreement is surprisingly good, especially in the bosonic sectors.
As has been previously discussed in \cite{Gross:1997mx, Kutasov:1993gq, Kutasov:1994xq,Gross:1995bp}, the existence of a topological sector in this theory, allows for a state of two-fermions (nominally bosonic) to also appear in the fermionic sector, for example.
The first two-particle threshold of $m^2=4m_{F_1}^2 = 22.9$ can be seen in all sectors.  In fact, we find it remarkable, that the lowest red-dashed trajectory seems to be matching the full diagonalization even at
quite small $\Delta_{max}$ (especially in the bosonic sectors), providing further evidence for effective conformal dominance.  Higher trajectories in the bosonic sectors match also, but only
once we go to higher $\Delta_{max}$.  This is due to the presence of the next two-fermion threshold at $m^2=(2.4+4.1)^2=42$, which complicates the mixing.  Consequently higher $\Delta_{max}$ is
required to properly resolve the spectrum.  For $\Delta_{max}=9$, in the bosonic sectors, all the states in the full theory match to the spectrum of the ``free two-particle model'' to better than $5\%$ accuracy
once both $|F_1\rangle \otimes |F_1\rangle$ and $|F_1\rangle \otimes |F_2\rangle$ thresholds are included.  We view this as evidence that we have correctly identified both thresholds using our
method.

The matching in the fermonic sectors is poorer, but evidence for the thresholds can be seen.  This can be do to the larger influence of the topological sector.  We do not seem to be
finding clear evidence for the existence of two-particle states, where one or both particles are bosons, in any sector.  It is possible, that for larger values of $\Delta_{max}$ the matching will improve
and two-particle states which include bosons will appear more clearly.

\section{Conclusion}
\label{conclusion}

In this paper we have explored the phenomena of effective conformal dominance.  This phenomena has been known for quite a while in the context of SUGRA backgrounds dual to confining gauge theories.
Indeed, in such backgrounds, all the lightest bound-states are fluctuations of the SUGRA fields, above the background, with the heavier states being described by stringy bulk modes
\cite{Klebanov:2000hb,Sakai:2004cn,Polchinski:2000uf}.  The SUGRA fields,
are, of course, dual to the lowest dimension primary operators of the gauge theory, the effective conformal sector (typically the stress tensor and its SUSY partners).  The stringy modes, on the other hand,
are dual to operators of high scaling-dimension, which are decoupled from the low-energy spectrum.  A natural question to ask is how robust is this phenomena of effective conformal dominance
over the low-energy spectrum?  In particular, does it hold only for theories where there is a hierarchy in the scaling-dimension of operators, or is it more universal?  Holographic models of 
strong dynamics of the AdS/QCD type, or more recently of the AdS/CMT type, would be on firmer ground if this dominance were more generic, especially if the ultimate goal is to use them
to understand physics quantitatively, and not just qualitatively.  In this context the central question becomes, if we ignore all bulk fields above a certain mass, or equivalently include only
operators whose dimension, $\Delta$, is bound by $\Delta_{max}$, what error should we expect on physical observables, such as the spectrum?  

In this paper, we attempted to answer the above questions in the context of a QCD$_{2A}$, a theory whose UV is a CFT without a hierarchy in the scaling-dimension of operators.  Nonetheless, 
we have found that effective conformal dominance holds, with high-scaling dimension operators decoupling exponentially fast, roughly as $e^{-\Delta_{max}}$, from the low-energy spectrum.  
This observation, has allowed us to suggest a new technique for solving this theory.  We compared our low-energy spectrum to numerical DLCQ results, and found very good agreement.
We find this to be an encouraging sign that effective conformal dominance could be a useful guide in obtaining a quantitative understanding of other strongly interacting systems.  It would 
be interesting to test this phenomena in other strongly coupled theories both in 2D, and in higher dimensions.

\section*{Acknowledgments}
We thank Siavosh R. Behbahani, Claudio Rebbi, Martin Schmaltz and specially Andy Cohen for useful discussions. We also thank Igor Klebanov for very helpful comments regarding an earlier
draft of the paper.  The work of EK and YX was supported in part by DOE grant DEFG02-01ER-40676 and NSF CAREER grant PHY-0645456. GMT acknowledges support from a DOE High Energy Physics Fellowship.

\appendix
\label{appendix}
\section{The quasi-primary operators}
\label{appquasi-primary}

In this appendix we describe in detail the construction of the conformal quasi-primary operators of multi-parton states. The Hamiltonian, which breaks the CFT with a single scale, is more easily diagonalized using the basis of these quasi-primary operators. We consider a single trace operator of adjoint fermions at a given dimension
\be
\mathcal{O}_{n+k/2}\equiv \frac{1}{N^{k/2}}\sum_{\sum s_i=n}c_{s_1, s_2, ..., s_k}\text{Tr}\left(\partial^{s_1}_{\mu_1}\psi_1\partial^{s_2}_{\mu_2}\psi_2...\partial^{s_k}_{\mu_k}\psi_k\right),
\label{primaries}
\ee
which, acting on the vacuum, creates a conformally symmetric $k$-fermion state.  Note that at $m=0$, because of the equation of motion $\partial_+\psi=0$, all the derivatives acting on the right moving state $\psi$ are with respect to the ``space-like'' coordinate $x^-$. Due to the chiral symmetry of the Hamiltonian, there is no mixing between the left and right moving states $\chi$ and $\psi$. In fact, it has been shown that \cite{Kutasov:1993gq,Gross:1997mx} even for the massive bound states of these fermions, the massive sector of $\chi$ only enters through the current $\bar{J}^{ab}=\chi^{ac}\chi^{cb}$, which is related to the right moving current $J$ by current conservation. Thus it is sufficient to consider in the quasi-primary operators only the right moving ones. The quasi-primary operator satisfies the commutation relation 
\be
[K^-, \mathcal{O}_{n+k/2}(x^-)]=i\left((x^-)^2\partial_-+x^-(2n+k)\right)\mathcal{O}_{n+k/2}(x^-). 
\ee
Here $K^-$ is the generator of the special conformal transformation along $x^-$. We will assume the superscript of $x^-$ implicit hereafter when there is no ambiguity.

The special conformal transformation generator $K$ that leaves the free fermion Lagrangian invariant is given by 
\be
K=-i\int dx  x^2\psi_{ij}\partial\psi_{ji}.
\ee
Thus the commutator of $K$ with a single trace operator in (\ref{primaries}) is\footnote{
Where we impose the anti-commutation relations at equal $x^+$: \begin{equation*} \{ \psi_{ij}(x^-),\, \psi_{kl}(y^-) \} = \tfrac{1}{2} \delta(x^- - y^-) \left( \delta_{il}\delta_{jk} -\tfrac{1}{N} \delta_{ij} \delta_{kl} \right). \end{equation*}
} 
\be
\begin{split}
&[K,  \text{Tr}\left(\partial^{s_1}\psi_1\partial^{s_2}\psi_2...\partial^{s_k}\psi_k\right)]\\
&=i\left(x \left(k+2\left( s_1+s_2+...+s_k\right)\right) + x^2 \partial \right) \text{Tr}\left(\partial^{s_1}\psi_1\partial^{s_2}\psi_2...\partial^{s_k}\psi_k\right) \\
&+i \sum_i s^2_i \text{Tr}\left(\partial^{s_1}\psi_1\partial^{s_2}\psi_2...\partial^{s_i-1}\psi_i ...\partial^{s_k}\psi_k\right)
\end{split}
\label{commutator}
\ee
In order to ensure $\mathcal{O}_{n+k/2}$ in (\ref{primaries}) is a quasi-primary operator we have to choose a set of $c_{s_1, s_2, ..., s_k}$  such that the second term in the equation above vanishes.

To proceed, we consider the $k$-fermion state in the momentum space using the mode expansion
\be
\psi_{ij}=\frac{1}{2\sqrt\pi}\int_0^\infty dp^+\left(b_{ij}(p^+)e^{-ip^+x^-}+b^\dag_{ji}(p^+)e^{ip^+x^-}\right)
\label{app_modeexpansion}
\ee
The coefficients and the derivatives of a quasi-primary operator acting on the vacuum can then be simplified into a homogeneous polynomial of fermion momenta, that is
\begin{align}
\label{quasi-primarystate}
&\mathcal{O}_{n+k/2}|0\rangle \nonumber \\
&=\int \left(\prod \frac{dp_i}{2\sqrt\pi}\right)\ \left(\sum_{\sum s_i=n} i^n c_{s_1, s_2, ..., s_k}p_1^{s_1} p_2^{s_2}...p_k^{s_k}\right)\frac{e^{i(p_1+p_2+...+p_k) x}}{N^{[k/2]}} \text{Tr}\left(b^\dag_1(p_1)...b^\dag_k(p_k)\right)|0\rangle. 
\end{align}
If we define the amplitude $f(p_1, p_2, ..., p_k)\equiv \left(\frac{1}{2\sqrt\pi}\right)^k\sum i^n \, c_{s_1, s_2, ..., s_k}p_1^{s_1} p_2^{s_2}...p_k^{s_k}$, then the homogeneity of $f$ requires that
\be
\sum_i p_i\frac{\partial}{\partial p_i} f=n f
\label{homogeneity}
\ee
whereas the vanishing of the second term in (\ref{commutator}) implies
\be
\sum_i \frac{\partial}{\partial p_i}\left( p_i\frac{\partial}{\partial p_i}f\right)=0.
\label{constraint}
\ee

When we diagonalize the Hamiltonian with these quasi-primary operators the total momentum $P$  of a state is fixed, $\sum p_i=P$. That means the function $f(p_1, p_2, ..., p_k)$ is defined on a simplex. This, together with eq. (\ref{homogeneity}), suggests that we can rewrite the function $f(p_1, p_2, ..., p_k)$ in terms of a set of angular variables defined by
\be
\begin{split}
p_{k}&=P \cos^2\theta_1,\\
p_{k-1}&=P \sin^2\theta_1 \cos^2\theta_2,\\
&...\\
p_{2}&=P \sin^2\theta_1 \sin^2\theta_2...\cos^2\theta_{k-1},\\
p_1&=P \sin^2\theta_1 \sin^2\theta_2...\sin^2\theta_{k-1}.\\
\end{split}
\label{replacement}
\ee
It is so arranged because it simplifies the structure of the Hamiltonian, as we will see in the next section. In the new variables eq.~(\ref{constraint}) becomes
\be
\begin{split}
&4 P^2 \frac{\di^2 f}{\di P^2}+4kP\dd{f}{P}+\frac{\di^2 f}{\di \theta_1^2}+\frac{2\left(k-2+\left(k-1\right)\cos2\theta_1\right)}{\sin 2\theta_1}\dd{f}{\theta_1}\\
&+\frac{1}{\sin^2\theta_1}\biggl(\frac{\di^2 f}{\di\theta_2^2}+\frac{2\left(k-3+\left(k-2\right)\cos2\theta_2\right)}{\sin 2\theta_2}\dd{f}{\theta_2}\\
&+\frac{1}{\sin^2\theta_2}\biggl(\frac{\di^2 f}{\di\theta_3^2}+\frac{2\left(k-4+\left(k-3\right)\cos2\theta_3\right)}{\sin 2\theta_3}\dd{f}{\theta_3}\\
&+...\\
&+\frac{1}{\sin^2\theta_{k-2}}\left(\frac{\di^2 f}{\di\theta_{k-1}^2}+\frac{2\cos2\theta_{k-1}}{\sin 2\theta_{k-1}}\dd{f}{\theta_{k-1}}
\right)\biggr)\biggr)... \biggr)=0
\end{split}
\label{constraintequation}
\ee
The first two terms reduce to $4n(n+k-1)f$ given that $f$ is a homogeneous function of order $n$. The variables of this equation can be separated and the solution is a product of  the well-known hypergeometric functions $_2F_1(a, b; a+b, \sin^2\theta_i)$, with $a$ and $b$ integers to be determined. Hence it is also equal to a product of Jacobi polynomials of the form $P^{(a+b-1,0)}_{-a}(\cos 2\theta_i)$. In general, the solution can be written as linear combinations of
\be
\begin{split}
&f_{n,l_1,l_2,...,l_{k-2}}\left(P, \theta_1, \theta_2...,\theta_{k-1}\right)\\
&=P^n (\sin\theta_1)^{^{2l_1}} (\sin\theta_2)^{2l_2} ... (\sin\theta_{k-2})^{2l_{k-2}}\\
&\times P^{(2l_1+k-2,0)}_{n-l_1}\left(\cos2\theta_1\right)P^{(2l_2+k-3,0)}_{l_1-l_2}\left(\cos2\theta_2\right)...P^{(2l_{k-2}+1,0)}_{l_{k-3}-l_{k-2}}\left(\cos2\theta_{k-2}\right)P_{l_{k-2}}\left(\cos2\theta_{k-1}\right),
\end{split}
\label{jacpsolution}
\ee
where $P_{l_{k-2}}\left(\cos2\theta_{k-1}\right)$ is a Legendre polynomial, a special case of the Jacobi polynomial. The indices $l_i$'s are integers that satisfy $n\geqslant l_1\geqslant l_2\geqslant l_3 ...\geqslant l_{k-2}\geqslant 0$. They parametrize the numbers of derivatives $\partial$'s acting on each of the $k$ fermion operators $\psi_i$. Because of the orthogonality of the Jacobi polynomials, and the fact that (\ref{jacpsolution}) is a solution to eq. (\ref{constraintequation}), $f_{n,l_1,l_2,...,l_{k-2}}$ are normalized up to a constant, and those with different indices are orthogonal.

Since the fermion operators in $\mathcal{O}_{n+k/2}$ are traced, the function $f$ in (\ref{quasi-primarystate})  has a well defined transformation property under a cyclic permutation of the momenta $p_i$,
\be
f(p_1,p_2,...,p_k) = (-1)^{k-1} f(p_2,p_3,...,p_k,p_1).
\label{cyclicsym}
\ee
Therefore by writing $f$ as a linear combination of  $f_{n,l_1,l_2,...,l_{k-2}}$ given in (\ref{jacpsolution}) with fixed $n$, there are additional constraints on the coefficients. These constraints can be read from the set of equations of $\theta_i$'s enforcing equality under the exchange of variables
\be
\begin{split}
\sin^2\theta_1&\rightarrow 1-\sin^2\theta_1\sin^2\theta_2...\sin^2\theta_{k-1}\\
\sin^2\theta_2 &\rightarrow \frac{\sin^2\theta_1\left(1-\sin^2\theta_2...\sin^2\theta_{k-1}\right)}{1-\sin^2\theta_1\sin^2\theta_2...\sin^2\theta_{k-1}}\\
\sin^2\theta_3 &\rightarrow \frac{\sin^2\theta_2\left(1-\sin^2\theta_3...\sin^2\theta_{k-1}\right)}{1-\sin^2\theta_1\sin^2\theta_3...\sin^2\theta_{k-1}}\\
&...\\
\sin^2\theta_{k-1} &\rightarrow \frac{\sin^2\theta_{k-2}\left(1-\sin^2\theta_{k-1}\right)}{1-\sin^2\theta_{k-2}\sin^2\theta_{k-1}}.\\
\end{split}
\ee
Although the cyclic transformation of the momenta in the angular variables $\theta_i$ is not as simple as that in the original momenta variables $p_i$, we will soon see an advantage of using these angular variables in parametrizing the Hamiltonian. The cancellation of the divergence becomes manifest, and a separation of the variables expedites the numerical calculation.

We also implement the T-parity symmetry on the quasi-primary operators, a symmetry of $\psi_{ij}\rightarrow\psi_{ji}$. In terms of the free fermion states, 
\be
|p_1,p_2,...,p_k \rangle = N^{-k/2} \text{Tr}(b^\dagger (p_1) b^\dagger (p_2) ... b^\dagger (p_k))| 0 \rangle,  \nonumber
\ee
this corresponds to
\be
\text{T} |p_1, p_2, ... , p_k \rangle = (-1)^\sigma |p_k, p_{k-1}, ... , p_1 \rangle,
\ee
where $\sigma = k/2$ for even $k$ and $\sigma = (k-1)/2$ for odd $k$. In the angular variables the transformation  $(p_1, p_2, ..., p_{k-1}, p_k)\rightarrow(p_k, p_{k-1}, ..., p_2, p_1)$  reads
\be
\sin^2\theta_i\rightarrow \frac{1-\prod^{k-i}_{j=1}\sin^2\theta_j }{1-\prod^{k-i+1}_{j=1}\sin^2\theta_j}.
\ee

It is straightforward to identify a quasi-primary operator through its momentum space amplitude $f$. For example, an operator (T-odd) with 4 partons and of dimension 4 has
\be
\begin{split}
\langle p_1, p_2, p_3, p_4 |\mathcal{O}_{2+4/2}|0\rangle\sim -1+2 \sin^2\theta_1+4\sin^2\theta_1\sin^2\theta_2-6\sin^4\theta_1\sin^2\theta_2\\
-4\sin^2\theta_1\sin^2\theta_2\sin^2\theta_3+6\sin^4\theta_1\sin^4\theta_2\sin^2\theta_3,
\end{split}
\ee
where we have used $\langle \theta_1, \theta_2, \theta_3|$ to denote $\langle p_1, p_2, ..., p_k|$ to make the dependance on the $\theta$'s more explicit.
It can be easily translated into 
\be
\begin{split}
\mathcal{O}_{2+4/2}&\sim\text{Tr}\biggl(\partial^2\psi_1\psi_2\psi_3\psi_4-\psi_1\partial^2\psi_2\psi_3\psi_4+\psi_1\psi_2\partial^2\psi_3\psi_4-\psi_1\psi_2\psi_3\partial^2\psi_4
\\
&-4\partial\psi_1\psi_2\partial\psi_3\psi_4+4\psi_1\partial\psi_2\psi_3\partial\psi_4\biggr)
\end{split}
\ee
by using eq.~(\ref{replacement}).
As another example, the quasi-primary operator in our basis for T-even with 3 partons and dimension 4.5 is
\be
\begin{split}
\mathcal{O}_{3+3/2}&\sim\text{Tr}\biggl(\partial^2\psi_1\partial\psi_2\psi_3-\partial\psi_1\partial^2\psi_2\psi_3-\partial^2\psi_1\psi_2\partial\psi_3
\\
&+\psi_1\partial^2\psi_2\partial\psi_3+\partial\psi_1\psi_2\partial^2\psi_3-\psi_1\partial\psi_2\partial^2\psi_3\biggr).
\end{split}
\ee

\section{The mass matrix in the angular variables}
\label{appmassmatr}

In this appendix we describe the calculation of the mass matrix elements in terms of the angular variables defined in eq.~(\ref{replacement}). The action for the model using light-cone coordinates, $x^{\pm} = (x^0 \pm x^1)/\sqrt{2}$, and choosing light-cone gauge $A_- = 0$, is
\be
S = \int dx^+ dx^- \text{Tr}\left( i \psi \partial_+ \psi + i \chi \partial_- \chi + \frac{1}{2 g^2} (\partial_- A_+)^2  + 2 A_+ \psi \psi  \right),
\label{action}
\ee
where $\psi$ and $\chi$ are respectively the right-mover and left-mover fermion (analogous to chiral fermions in 4 dimensions). It is convenient to choose $x^+$ as the time component, since with this choice $\chi$ and $A_+$ are not propagating degrees of freedom. Integrating out this non-propagating fields one finds that the momentum operators are
\be
\begin{split}
P^+ & = \int dx^- \text{Tr} \left( i \psi \partial_- \psi \right), \\
P^- & = \int dx^- \text{Tr} \left(  -2 g^2 \psi^2 \frac{1}{\partial^2_-} \psi^2 \right).
\end{split}
\label{momentum operator}
\ee

The mass matrix is given by $(M^2)_{ij} = \langle \mathcal{O}_i | 2 P^+ P^- | \mathcal{O}_j \rangle$. In order to compute this matrix it is convenient to first write the matrix element 
\begin{align}
\langle p_1, p_2, ... , p_k | & 2 P^+ P^-  | \psi \rangle = \frac{ g^2 N}{\pi (x_1 +x_2)^2} \int_0^{x_1+x_2} dy \psi_k (y, x_1+x_2 - y, x_3, ... , x_k) \nonumber \\
								& + \frac{g^2 N}{\pi}  \int_0^{x_1+x_2} \frac{dy}{(x_1-y)^2} \left[  \psi_k(x_1, x_2, x_3,..., x_k) - \psi_k (y, x_1+x_2 -y,x_3,...,x_k)\right] \nonumber \\
								& + \frac{g^2 N}{\pi} \int_0^{x_1} dy \int_0^{x_1-y} dz \psi_{k+2}(y,z,x_1-y-z,x2,...,x_k)\left[ \frac{1}{(y+z)^2}-\frac{1}{(x_1-y)^2} \right] \nonumber \\
								& + \frac{g^2 N}{\pi} \psi_{k-2}(x_1+x_2+x_3,x_4,...,x_k)\left[ \frac{1}{(x_1+x_2)^2}-\frac{1}{(x_2+x_3)^2} \right] \nonumber \\
								& \pm \text{cyclic permutations of } (x_1, x_2, ..., x_k)
\end{align}
where $x_i = p_i/\sum p_j$ and $\psi_k (x_1,x_2,...,x_k) = \langle p_1, p_2, ...,p_k | \psi \rangle$. The sign in the last term is always positive for odd $k$ and alternates with each cyclic permutation for even $k$. 

Introducing a complete set of gauge-singlet states $\sum_k \, k^{-1} | p_1, p_2, ...,p_k \rangle  \langle p_1, p_2, ...,p_k | $, using the cyclic properties of the wave-functions and changing to the angular variables of eq.~(\ref{replacement}), one can show that
\begin{align}
\frac {[M^2]_{i, j}}{g^2 N/ \pi} = & \int  \prod^{k-1}_{m=1} \sin^{2(k-m-1)}  \theta_m d\sin^2\theta_m \left[ \int d\sin^2 \phi \left( \, \frac{\chi^*_k (\theta_1,...,\theta_{k-1}) \, \psi_k(\theta_1,..., \theta_{k-2},\phi)}{\sin^2 \theta_1 \sin^2 \theta_2 .... \sin^2 \theta_{k-2}} \,   \right. \right. \nonumber \\ 
&+\, \left. \left. \frac{1}{2} \frac{\left( \chi_k (\theta_1, ..., \theta_{k-1}) - \chi_k(\theta_1, ..., \theta_{k-2}, \phi) \right)^* \left(\psi_k (\theta_1, ..., \theta_{k-1}) - \psi_k(\theta_1, ..., \theta_{k-2}, \phi)  \right)}{\sin^2 \theta_1 ... \sin^2 \theta_{k-2} (\sin^2 \theta_{k-1}-\sin^2 \phi)^2}    \right) \right. \nonumber \\
 & + \left.  \int d\sin^2 \phi_1 d\sin^2 \phi_2 \,   \chi^*_k(\theta_1,..., \theta_{k-1}) \left(\frac{\psi_{k+2}(\theta_1, ... , \theta_{k-1}, \phi_1, \phi_2) - \tilde \psi_{k+2}}{\sin^2 \phi_1} \right)  \right. \nonumber \\
 & + \left.   \left( \frac{ \chi^*_k (\theta_1, ..., \theta_{k-1}) - \tilde \chi^*_k  }{\sin^4\theta_1 ... \sin^4 \theta_{k-2} }   \right)   \psi_{k-2}(\theta_1, ..., \theta_{k-3}) \right].
\label{massmatrixangles}
\end{align}
All the angles are integrated from 0 to $\pi/2$. Here $\psi_k(\theta_1, ..., \theta_{k-1}) = \langle p_1, ..., p_k | \mathcal{O}_j \rangle$ and $\chi_k(\theta_1, ..., \theta_{k-1}) = \langle p_1, ..., p_k | \mathcal{O}_i \rangle$, and we have defined
\be
\begin{split}
\tilde \psi_k(\theta_1, ..., \theta_{k-1}) & = \psi_k (\theta_1, ..., \theta_{k-3}, \tilde \theta_{k-2}, \tilde \theta_{k-1}), \\
\sin^2 \tilde \theta_{k-2} & = 1- \sin^2 \theta_{k-2} \sin^2 \theta_{k-1} \, , \\  \sin^2 \tilde \theta_{k-1} & = \frac{1- \sin^2 \theta_{k-2}}{1- \sin^2 \theta_{k-2} \sin^2 \theta_{k-1}} \, ,
\end{split}
\ee
with $\tilde \chi$ defined by replacing the last two angles of $\chi$ in the same way as above. The transformation $\theta\rightarrow\tilde\theta$ leads to an exchange of the parton momenta $p_1\leftrightarrow p_3$.

Note that the first two terms on the right hand side of ~(\ref{massmatrixangles}) become products of $k-1$ one-variable integrals. This significantly reduces the difficulties in numerical computation, which otherwise would be a $k$-dimensional integral evaluated on a simplex. In addition, all the potential divergences from the first two terms now explicitly cancel. 

In order to see that there is no divergence in the last two terms, further inspection is required. Focusing on the last term, the cancelation of divergences can be made explicit by adding an extra term that vanishes upon integrating out all variables. This is achieved by replacing $\psi_{k-2}$ in the last term by $\psi_{k-2}-\psi_{k-2}\big|_{p_1 = p_2 =0}$, to explicitly cancel the divergence at $p_1+p_2=0$. In terms of the original momentum variables $(p_1,...,p_k)$, one can easily show that the contribution from the $\psi_{k-2}\big|_{p_1 = p_2 =0}$ term vanishes by using the cyclicity of the wave-function, and thus does not contribute to the final result. Here there is a subtlety, that when we set $p_1=p_2 = 0$, we have to replace the total momentum $P$ in ~(\ref{jacpsolution}) by the sum of the remaining momenta $\sum^k_{i=3} p_i$ and relate the new angles $\theta^\prime_i (i=1,2,...,k-3)$, associated with the variables $(p_3, p_4, ...,p_k)$,  to the original angles $\theta_i (i=1,2,...,k-1)$. Again it is not hard to work out the transformation. Once all the divergences are explicitly canceled we are left with multi-dimensional polynomials to integrate instead of ratios of polynomials. Thus by writing the mass matrix in the angular variables, we are able to  expedite the calculation and avoid issues with the convergence of numerical integration near singularities. 

On the other hand, we currently lack an optimized strategy for performing the algebraic manipulations required to construct  $\psi_{k-2}\big|_{p_1 = p_2 =0}$ and for explicitly canceling the $1/\sin^2 \! \theta_{k-2}$ pole. The addition of the $\psi_{k-2}\big|_{p_1 = p_2 =0}$ term also spoils the factorizability of the integrals and forces us to deal with multi-dimensional integrals over the angles. This constitutes one of the main bottle-necks for extending our computations to larger $\Delta_{max}$.

\section{Quasi-primary operators in the effective two free-particle models}
\label{freetwopartbasis}

In this appendix we present the general case for the quasi-primary operators of two free particles, where the particles can be two fermions, a boson and a fermion or 2 bosons. The general form of quasi-primary operators bilinear in the fields can be written in terms of Jacobi Polynomials~(see, e.g., \cite{Braun:2003rp})
\be
\mathcal{O}_{\Delta_n} = \left\{
\begin{array}{ll}
\psi_1 (x) P_n^{(0,0)} \! \left(  \overleftarrow\partial - \overrightarrow \partial \right) \psi_2(x) \, , & \text{for 2 fermions}, \\
 \partial \phi (x) P_n^{(1,0)} \! \left( \overleftarrow \partial - \overrightarrow \partial \right) \psi(x) \, , & \text{for a boson and a fermion} \\
 \partial \phi_1 (x) P_n^{(1,1)} \! \left(  \overleftarrow \partial - \overrightarrow \partial \right) \partial \phi_2(x)\, , & \text{for 2 bosons}  .
\end{array} \right.
\label{freepartoperators}
\ee
Here the dimension of the operator is $\Delta = n + d_1 + d_2$, where $d_i = 1/2$ for fermions and $d_i = 1$ for bosons.

The expansion of the boson field in terms of creation and annihilation operators is given by
\be
\phi(x^-) = \frac{1}{\sqrt{2 \pi}} \int_0^\infty \frac{dp^+}{\sqrt{2 p^+}} \left( a(p^+) e^{-ip^+ x^-} + a^\dag (p^+) e^{ip^+ x^-}  \right).
\ee
It contains an extra factor of $\sqrt{p^+}$ compared to the fermion operator in~eq.~(\ref{app_modeexpansion}). Hence the wave-function of a state created by bosons has extra factors of square root of momenta, compared to the fermion case, in addition to the difference in polynomials discussed in the previous paragraph.

The effect of a topological sector is simply accounted for, by adding to the operator dimension $\Delta$ an extra factor of $1/2$ or $-1/2$, which is the dimensional difference between a bosonic single-particle state and a fermionic single-particle state. For example, the two-fermion threshold of the state $F_1\otimes F_1$ appears in the fermionic sector. In order to match the QCD$_{2A}$ spectrum with the free $F_1\otimes F_1$ spectrum generated by two fermion operators, we subtract $1/2$ from the operator dimension $\Delta$, assuming, from the point of view of operator dimension,  that the topological sector converts one of the fermions into a boson.

\bibliography{2dqcdref}{}
\bibliographystyle{JHEP}

\end{document}